%% file: maxij0556_sugizaki.tex
\documentclass{pasj00}   
\usepackage{enumerate}

\usepackage{color}

\newcommand{\riken}{1} 
\newcommand{\agu}{2} 
\newcommand{\jaxa}{3} 
\newcommand{\psu}{4} 
\newcommand{\kyoto}{5} 
\newcommand{\sokendai}{6} 
\newcommand{\isas}{7} 
\newcommand{\tit}{8} 
\newcommand{\osaka}{9} 
\newcommand{\chuo}{10} 
\newcommand{\waseda}{11} 
\newcommand{\nichidaishi}{12} 
\newcommand{\nichidai}{13} 
\newcommand{\miyazaki}{14} 

\newcommand{\ledd}{L_{\rm EDD}}

\begin{document}
\SetRunningHead{Sugizaki et al.}{Spectral Evolution of MAXI J0556$-$332}


\title{Spectral Evolution of a New X-ray Transient MAXI J0556$-$332
  Observed by MAXI, Swift, and RXTE}

\author{
Mutsumi \textsc{Sugizaki}\altaffilmark{\riken},
Kazutaka \textsc{Yamaoka}\altaffilmark{\agu},
Masaru \textsc{Matsuoka}\altaffilmark{\riken,\jaxa},
Jamie A. \textsc{Kennea}\altaffilmark{\psu},
Tatehiro \textsc{Mihara}\altaffilmark{\riken},
%
Kazuo \textsc{Hiroi}\altaffilmark{\kyoto},
Masaki \textsc{Ishikawa}\altaffilmark{\sokendai},
Naoki \textsc{Isobe}\altaffilmark{\isas},
Nobuyuki \textsc{Kawai}\altaffilmark{\tit},
Masashi \textsc{Kimura}\altaffilmark{\osaka},
Hiroki \textsc{Kitayama}\altaffilmark{\osaka},
Mitsuhiro \textsc{Kohama}\altaffilmark{\jaxa},
Takanori \textsc{Matsumura}\altaffilmark{\chuo},
Mikio \textsc{Morii}\altaffilmark{\tit},
Yujin E. \textsc{Nakagawa}\altaffilmark{\waseda},
Satoshi \textsc{Nakahira}\altaffilmark{\riken},
Motoki \textsc{Nakajima}\altaffilmark{\nichidaishi},
Hitoshi \textsc{Negoro}\altaffilmark{\nichidai},
Motoko \textsc{Serino}\altaffilmark{\riken},
Megumi \textsc{Shidatsu}\altaffilmark{\kyoto},
Tetsuya \textsc{Sootome}\altaffilmark{\riken},
Kousuke \textsc{Sugimori}\altaffilmark{\tit},
Fumitoshi \textsc{Suwa}\altaffilmark{\nichidai},
Takahiro \textsc{Toizumi}\altaffilmark{\tit},
Hiroshi \textsc{Tomida}\altaffilmark{\jaxa},
Yoko \textsc{Tsuboi}\altaffilmark{\chuo},
Hiroshi \textsc{Tsunemi}\altaffilmark{\osaka},
Yoshihiro \textsc{Ueda}\altaffilmark{\kyoto},
Shiro \textsc{Ueno}\altaffilmark{\kyoto},
Ryuichi \textsc{Usui}\altaffilmark{\tit},
Takayuki \textsc{Yamamoto}\altaffilmark{\riken},
Makoto \textsc{Yamauchi}\altaffilmark{\miyazaki},
Kyohei \textsc{Yamazaki}\altaffilmark{\chuo},
Atsumasa \textsc{Yoshida}\altaffilmark{\agu},
and the MAXI team}

\altaffiltext{\riken}{MAXI team, Institute of Physical and Chemical
  Research (RIKEN), 2-1 Hirosawa, Wako, Saitama 351-0198}
\email{sugizaki@riken.jp}

\altaffiltext{\agu}{Department of Physics and Mathematics, Aoyama Gakuin
  University, 5-10-1 Fuchinobe, Sagamihara, Kanagawa 229-8558}

\altaffiltext{\jaxa}{ISS Science Project Office, Institute of Space
  and Astronautical Science (ISAS), Japan Aerospace Exploration Agency
  (JAXA), 2-1-1 Sengen, Tsukuba, Ibaraki 305-8505}

\altaffiltext{\psu}{ Department of Astronomy and Astrophysics, 0525 Davey
  Laboratory, Pennsylvania State University, University Park, PA 16802, USA}

\altaffiltext{\kyoto}{Department of Astronomy, Kyoto University,
  Oiwake-cho, Sakyo-ku, Kyoto 606-8502}

\altaffiltext{\sokendai}{School of Physical Science, Space and Astronautical
  Science, The Graduate University for Advanced Studies (Sokendai),
  3-1-1 Yoshinodai, Chuo-ku, Sagamihara, Kanagawa 252-5210}

\altaffiltext{\isas}{Institute of Space and Astronautical Science(ISAS),
  Japan Aerospace Exploration Agency(JAXA) ,3-1-1 Yoshino-dai,
  Chuo-ku, Sagamihara, Kanagawa 252-5210}

\altaffiltext{\tit}{Department of Physics, Tokyo Institute of Technology,
  2-12-1 Ookayama, Meguro-ku, Tokyo 152-8551}

\altaffiltext{\osaka}{Department of Earth and Space Science, Osaka
  University, 1-1 Machikaneyama, Toyonaka, Osaka 560-0043}

\altaffiltext{\chuo}{Department of Physics, Chuo University, 1-13-27 Kasuga, Bunkyo-ku, Tokyo 112-8551}

\altaffiltext{\waseda}{Research Institute for Science and Engineering,
  Waseda University, 17 Kikui-cho, Shinjuku-ku, Tokyo 162-0044}

\altaffiltext{\nichidaishi}{School of Dentistry at Matsudo, Nihon University, 2-870-1
Sakaecho-nishi, Matsudo, Chiba 101-8308}

\altaffiltext{\nichidai}{Department of Physics, Nihon University, 1-8-14,
  Kanda-Surugadai, Chiyoda-ku, Tokyo 101-8308}

\altaffiltext{\miyazaki}{Department of Applied Physics, University of
  Miyazaki, 1-1 Gakuen Kibanadai-nishi, Miyazaki, Miyazaki 889-2192}




%


\maketitle

\begin{abstract}

  We report on the spectral evolution of a new X-ray transient, MAXI
  J0556$-$332, observed by MAXI, Swift, and RXTE. The source was
  discovered on 2011 January 11 (MJD$=$55572) by MAXI Gas Slit Camera
  all-sky survey at $(l,b)=(238^\circ.9, -25^\circ.2)$, relatively away
  from the Galactic plane.  Swift/XRT follow-up observations
  identified it with a previously uncatalogued bright X-ray source and
  led to optical identification.  For more than one year since its
  appearance, MAXI J0556$-$332 has been X-ray active, with a 2-10 keV
  intensity above 30 mCrab.  The MAXI/GSC data revealed rapid X-ray
  brightening in the first five days, and a hard-to-soft transition in
  the meantime.  For the following $\sim 70$\,days, the 0.5--30 keV
  spectra, obtained by the Swift/XRT and the RXTE/PCA on an almost
  daily basis, show a gradual hardening, with large flux variability.
  These spectra are approximated by a cutoff power-law with a photon
  index of 0.4--1 and a high-energy exponential cutoff at 1.5--5\,keV,
  throughout the initial 10 months where the spectral evolution is
  mainly represented by a change of the cutoff energy.  To be more
  physical, the spectra are consistently explained by thermal emission
  from an accretion disk plus a Comptonized emission from a boundary
  layer around a neutron star. This supports the source identification
  as a neutron-star X-ray binary.  The obtained spectral parameters
  agree with those of neutron-star X-ray binaries in the soft state,
  whose luminosity is higher than $1.8\times 10^{37}$ erg s$^{-1}$.
  This suggests a source distance of $>17$ kpc.


\end{abstract}

\section{Introduction}
\label{sec:intro}

Galactic X-ray binaries distributed near the Galactic plane,
exhibiting some of the brightest X-ray fluxes in the sky, have been
well studied since the beginning of X-ray astronomy
(e.g. \cite{1981SSRv...29..221H}).
Their X-ray emission is thought to occur when the gas from a companion
(often late type) star accretes onto a compact star, a neutron star
(NS) or a black hole (BH).  They often show a large degree of X-ray
flux variability and long periods of quiescence, only appearing in
single (or sometimes recurring) periods of transient X-ray activity.
Many attempts have been made to understand the behavior of these
objects in a unified scheme, particularly employing accretion-disk
theory \citep{1973A&A....24..337S}.

So far, the standard-disk picture has successfully explained the X-ray
emission from NS binaries with weak magnetic fields in their bright
phase (\cite{mitsuda1984PASJ}; \cite{1986ApJ...308..635M}).  Our next
task is to understand their variations, particluarly spectral state
transitions that are often seen when these sources exhibit transient
outbursts.
Although RXTE, INTEGRAL, and Swift survey observations with wide sky
coverages provided useful information, few works have been done on the
initial transitions.  In particular, these studies on NS X-ray
transients are limited because they are much fainter than BH X-ray
novae.

The unbiased all-sky monitoring with Monitor of All X-ray Image (MAXI;
\cite{matsuoka_pasj2009}) allows us to detect X-ray novae and transients, 
and to follow their intensity evolution from the beginning to the end.
The MAXI mission started in 2009 August and has already detected
several X-ray transients and novae in their initial phase
(e.g. \cite{2010PASJ...62L..27N}; \cite{2011arXiv1110.6512Y}; MAXI web
site\footnote{http://maxi.riken.jp/top/}).
\citet{asai_lmxb} studied the initial rising behavior of outbursts
from two transient NS low-mass X-ray binaries (LMXBs), Aql X-1 and 4U
1608-52, and then derived a relation between the initial hard-state
duration and the hard-to-soft transition luminosity.

In the constellation Columba, a new X-ray transient, MAXI J0556$-$332,
was discovered by the MAXI Gas Slit Camera (GSC; \cite{mihara_gsc1})
at 9:21 (UT) on 2011 January 11 \citep{2011ATel.3102....1M}.  Its
position in Galactic coordinates, ($l$,$b$)=(238$^\circ$.9,
-25$^\circ$.1), is relatively away from the Galactic plane.  A Swift
\citep{2004ApJ...611.1005G} follow-up observation confirmed a bright
uncatalogued X-ray source within the MAXI error circle and localized
the source position at J2000 coordidates of ($\alpha$, $\delta$) =
(89$^\circ$.19300, -33$^\circ$.17451) = (\timeform{5h56m46s.32},
\timeform{-33D10'28''.2}) with the positional uncertainty of
\timeform{1''.7} \citep{2011ATel.3103....1K}.  The X-ray source
agrees in position with an optical star with a $B$-magnitude of 19.4.
RXTE Target-of-Opportunity (ToO) observations were also performed.
The results revealed complex time variability, together with energy
spectra that can be represented by a sum of a multi-color disk
blackbody ({\tt diskBB} in Xspec terminology) and a blackbody ({\tt
  BB}) (\cite{2011ATel.3106....1S}; \cite{2011ATel.3110....1S};
\cite{2011ATel.3112....1B}).
Based on color-color (CD) and hardness-intensity (HID) diagrams
extracted from the RXTE data, \citet{2011ATel.3650....1H} suggested
that the source is a transient neutron-star Z source.  They estimated
the source distance to be 20--35 kpc, from a change of the CD/HID
tracks which is thought to occur at the same luminosity as in another
transient Z source, XTE J1701$-$462 (e.g. \cite{2010ApJ...719..201H}).

Follow-up observations of the optical counterpart found that the star
had brightened to $R \sim 17.8$ from its USNO-B1.0 magnitude of 
$R=19.9$, and the spectrum
revealed emission lines 
indicating the presence of an accretion disk around the compact
accretor \citep{2011ATel.3104....1H}.  It continued to brighten in the $R$-band 
until January 17, but slightly faded on January 18
\citep{2011ATel.3116....1R}.  
Optical spectroscopy
revealed narrow emission lines in the Bowen blend, and a period search
in radial velocities of these lines provided two candidate orbital
periods, $16.43\pm 0.12$ hrs and $9.754\pm 0.048$ hrs
\citep{2011arXiv1111.6946C}.
A radio observation on January 19 
with the Australia Telescope Compact
Array detected a faint radio source 
\citep{2011ATel.3119....1C}.  Both the
optical-to-X-ray and the radio-to-X-ray flux ratios suggest that the
X-ray source is probably a NS binary rather than a BH binary, if it
belongs to our Galaxy with a distance less than 20 kpc
(\cite{2011ATel.3116....1R}; \cite{2011ATel.3119....1C}).
An XMM-Newtona observation with the reflection grating spectrometer
(RGS) detected a strong emission line near 24.8 \AA whose center
energy is consistent with the Ly$\alpha$ transition of N VII in the
rest frame.  From an extremely high N/O line ratio revealed by this
observation, the donor star is suggested to be a peculiar exotic star
such as a hot subdwarf (sdB, sdO) or a white dwarf
\citep{2011ApJ...743L..11M}.

These observational results suggest that MAXI J0556$-$332 is an X-ray
binary, involving a collapsed object, located in the Galactic halo.  
The companion is not a regular late-type star, so that 
the source distance has not been determined better than the
X-ray estimate of 20--35 kpc by \citet{2011ATel.3650....1H}. 
Most of these results favor the collapsed component being a NS,
although the BH scenario has not yet been completely ruled out.

This paper presents the X-ray behavior of MAXI J0556$-$332, including
an initial transition and continuous long-term variations, observed by
the MAXI GSC. We also analyzed spectral variations using data taken by
the Swift/X-Ray Telescope (XRT; \cite{2005SSRv..120..165B}) and the
RXTE/Proportional Counter Array (PCA; \cite{Jahoda2006}) on an
approximately daily cadence. We describe the observations and the data
reduction in section \ref{sec:obs_data}, and present the analysis
results in section \ref{sec:results}.  In section
\ref{sec:discussion}, the origin of the X-ray emission and its
evolution are discussed.  All the quoted errors are hereafter given at
the 90\% confidence limit, unless otherwise specified.

\section{Observation and Data Reduction}
\label{sec:obs_data}

\subsection{MAXI/GSC} 

Every 92-minute orbital revolution,
MAXI on the International Space Station (ISS) scans almost the whole sky
with two kinds of X-ray cameras: the GSC
working in the 2--30 keV energy band,
and the Solid-state Slit Camera (SSC; \cite{tsunemi_ssc2};
\cite{tomida_ssc1}) in the 0.5--10 keV band.
The new X-ray transient, MAXI J0556$-$332, first detected by the GSC
on 2011 January 11, brightened to 80 mCrab in the 4--10 keV band
within a day \citep{2011ATel.3102....1M}.  The upper limit on the
average 4--10 keV flux prior to the detection is 1.2 mCrab $=1.5\times
10^{-11}$ erg cm$^{-2}$ s $^{-1}$ \citep{2011PASJ...63S.677H}.
The source was also detected by the SSC when it was discovered by the
GSC.  However, the time coverage of the SSC was too limited to study
spectral and flux changes.  We therefore concentrate on the GSC data.

Reduction and analysis of the GSC data were carried out following the standard
procedure described by \citet{sugizaki_gsc2}.  The source event data
were extracted from a rectangular area of $3^\circ.6$ along the
detector anode wires and $3^\circ.0$ in the scan direction centered at
the source position; the latter corresponds to the point spread
function for each single scan transit of $\sim 40$ seconds.  The
background was collected from data taken in the same anode area just
before and after each scan transit.  
With the same analysis procedure, we processed the light curve of the
Crab nebula for the same period, and then confirmed that 
calibration uncertainties in the standard light curves in the 2--4
keV, 4--10 keV, and 10--20 keV energy bands are at most 5\% at the
1-$\sigma$ level.

\begin{figure}
\begin{center}
\FigureFile(8.6cm,){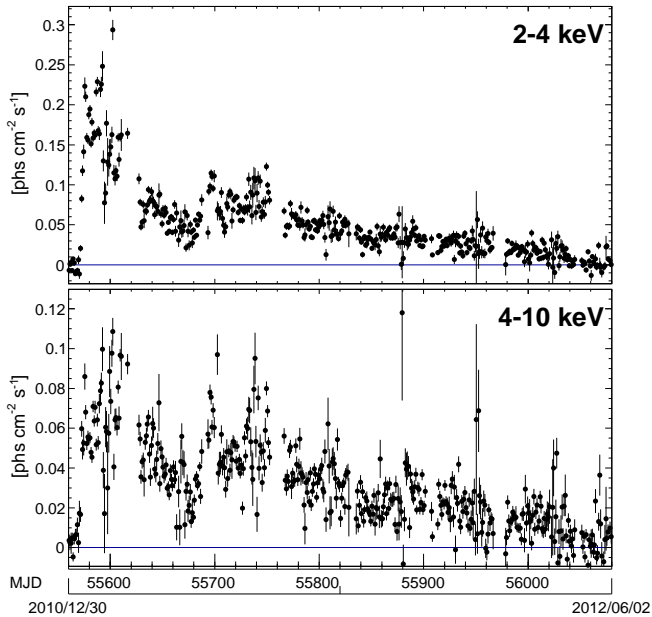}
\end{center}
\caption{ MAXI/GSC light curves of MAXI J0556$-$332 in 2-4 keV (top)
  and 4-10 keV (bottom).  Each data point represents a daily average.
}
\label{fig:gsc_lc}

\end{figure}

Figure \ref{fig:gsc_lc} shows the obtained GSC light curve of MAXI
J0556$-$332 in the 2--4 keV and 4--10 keV bands for the entire
1.5-year active period. These data represent the unfolded photon flux
per 1-day time bin.  In the 10--20 keV band, the GSC did not detect
any significant flux above 3-$\sigma$ confidence limit, 
which is typically 0.015 photons cm$^{-2}$ s$^{-1}$ (45 mCrab).

\subsection{Swift/XRT} 

Swift/XRT pointing observations of MAXI J0556$-$332 were performed
over 112 epochs until 2011 November 22 utilizing the Windowed Timing
(WT) mode and and a typical exposure of 0.5--1 ks.  Using the archival
Swift/XRT data, we investigated the evolution of the 0.3--10 keV 
energy spectrum.
The data reduction and analysis were performed using the Swift
analysis software version 3.8, released as a part of HEASOFT 6.11 and
CALDB (calibration database) files of version 20110915, provided via
NASA/GSFC.
Since the WT data are 1-dimensional, only spatial information in the
CCD detector X (DETX) direction is available. The source events were
collected from a 40 pixel wide region centered on the target position,
and the backgrounds were collected from a region with the same width
as the source region and 40-pixel away from the target along the DETX
direction.  In spectral model fitting, we used an XRT response matrix
file, {\tt swxwt0to2s6\_20010101v014.rmf}, and ancillary response
files built by {\tt xrtmkarf}.  A systematic error of 2\% was
implemented \citep{2009A&A...494..775G}.

\subsection{RXTE/PCA}

From 2011 January 13 to 2011 December 29, MAXI J0556$-$332 was obseved
with RXTE/PCA on an almost daily basis with a typical exposure of 1--2
ks. The obtained data provides useful information in the energy range
from 3 to 30 keV.
We performed reduction of the PCA data with the standard RXTE analysis
tools released as a part of HEASOFT 6.11 and the CALDB files of
version 20111102 provided via NASA/GSFC.  We used the PCA {\tt
  standard-2} data with a time resolution of 16-s for the spectral
analysis, and the {\tt Good-Xenon} data with a time resolution of
$1$-$\mu$s for the light-curve analysis.  All the data was screened
with the standard selection criteria: the spacecraft pointing
offset is smaller than $0^\circ.02$, the earth-limb elevation angle is
larger than $10^\circ$, and the time since the last South Atlantic
Anomaly passage is longer than 30 minutes.  We used event data
detected only on the top layer of the Proportional Counter Unit (PCU)
\#2, which is the best calibrated among all counter units. The
background was estimated using the archived background model provided
by the instrument
team\footnote{http://heasarc.gsfc.nasa.gov/docs/xte/pca\_news.html}.
The response matrix files were built with {\tt pcarsp} for each
pointing observation.  A systematic error of 0.5\% was applied.

\section{Analysis and Results}
\label{sec:results}

\subsection{Long-Term Light Curve and Color Variations}
\label{sec:longlc}

\begin{figure*}
\begin{center}
\FigureFile(13.cm,){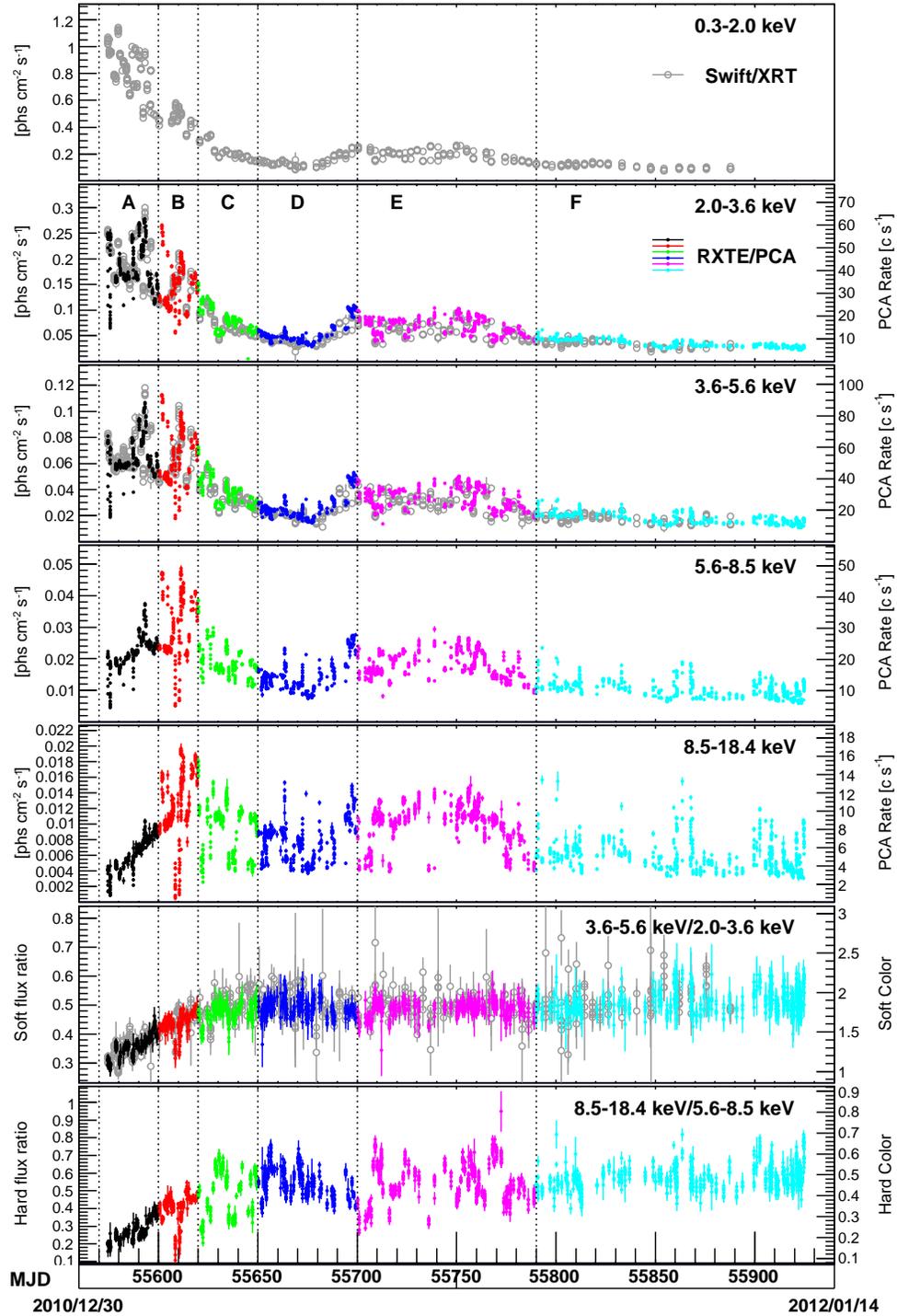}
\end{center}
\caption{ 

  Light curves in 0.3--2.0 keV, 2.0--3.6 keV, 3.6--5.6 keV, 5.6--8.5
  keV, and 8.5-18.4 keV energy bands obtained by the Swift/XRT (gray)
  and the RXTE/PCA.  Black, red, green, blue, purple and cyan specify
  intervals A, B, C, D, E, and F respectively, as labeled in the
  second panel.  Soft color (3.5--5.6 keV / 2.0--3.6 keV) and hard
  color (8.5--18.4 keV / 5.6--8.5 keV) of the RXTE/PCA data are shown
  in the bottom two panels.  All panels utilize 256-s time bin.  All
  the vertical error bars represent the 1-$\sigma$ statistical
  uncertainty.

}
\label{fig:lc_hr}
\end{figure*}

In order to investigate long-term flux and hardness variations with a
good statistic accuracy, we first extracted five-band X-ray light
curves from the Swift/XRT (0.3--6 keV) and RXTE/PCA (2--20 keV) data.
Figure \ref{fig:lc_hr} shows the results from 2011 January 13
(MJD$=$55574) to 2011 December 29 (MJD$=$55924), together with
soft-color (3.6--5.6 keV / 2.0-3.6 keV) and hard-color (8.5--18.4 keV
/ 5.6--8.5 keV) variations.  In order to enable direct comparison
among the different instruments, the observed count rates per 256-s
time bin in each instrument were converted to the unfolded photon
flux, assuming that the spectrum has a power-law with a photon index
of 1 and a high-energy exponential cutoff at $E_{\rm cut}=$ 3 keV,
with an interstellar absorption of $0.29\times 10^{21}$ cm$^{-2}$;
these are based on the spectral analysis in section
\ref{sec:ana_spec}.  The Swift/XRT and RXTE/PCA data in the same
energy band, taken within about a day, thus mostly agree with each
other, which confirms general consistency between the two instruments.
The large flux discrepancy seen for a few data points close in time is
indicative of flux changes between the pair of observations, which
were not exactly simultaneous.

These light curves show complex energy-dependent variations,
particularly during the initial $\sim$ 50 days.  In the lowest energy
band (0.3--2 keV), the flux reached the maximum within a few days
after the source emergence, and then decayed in about 50 days. In the
highest energy band (8.5--18.4 keV), the flux in contrast increased
gradually, reaching the maximum with large short-time variability on
about the 50th day.  The contrast between the soft and hard flux
evolutions are clearly reflected in the soft-color and hard-color
variations, because both colors largely increased during the initial
50 days.

As shown in figure \ref{fig:lc_hr}, we divided the entire
observation period into six intervals, A, B, C, D, E, and F; they are
characterized as follows.
Int-A: initial (30 days) phase when the intensity below 2 keV was highest 
and the hardest-band intensity increased gradually.
Int-B: brightest phase in the 5.6-8.5 keV and harder bands, with large variability in the whole band. 
Int-C: decaying phase where the large variability in the 5.6-8.5 keV and harder bands still remains.
Int-D: intermission phase where the average flux is lower than in the intervals 
just before and after.
Int-E: re-brightening phase in the whole band with moderate variability.
Int-F: low activity phase with a similar average flux to Int-D.

\begin{figure*}
\begin{minipage}[t]{9.4cm}
\FigureFile(9.4cm,){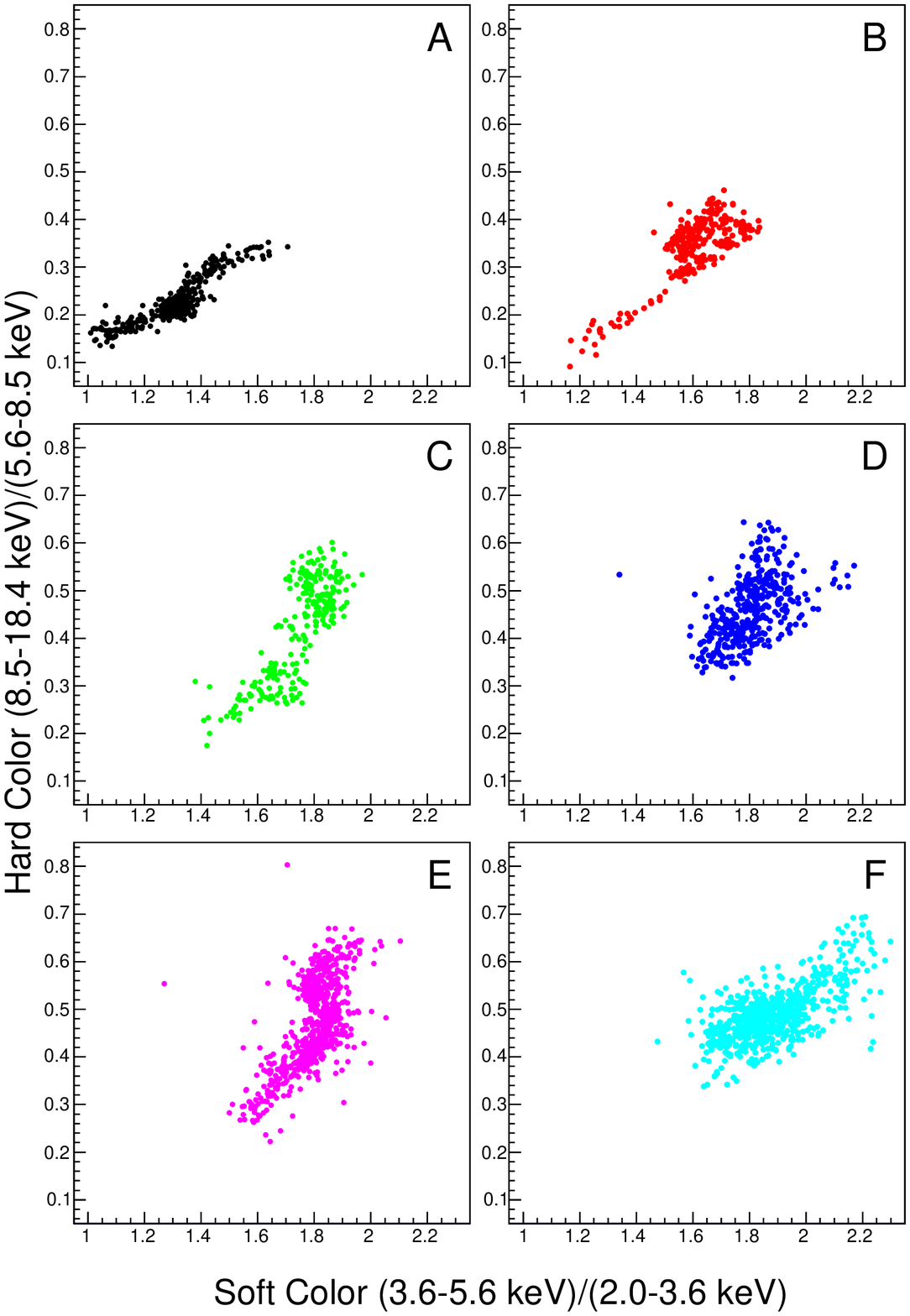}
\caption{ Color-color diagrams (CDs) extracted from the RXTE-PCA data,
  for individual intervals of A, B, C, D, E, and F defined in figure
  \ref{fig:lc_hr}.  }
\label{fig:ccdigaram}
\end{minipage}
\hspace{5mm}
\begin{minipage}[t]{7.6cm}
\FigureFile(7.6cm,){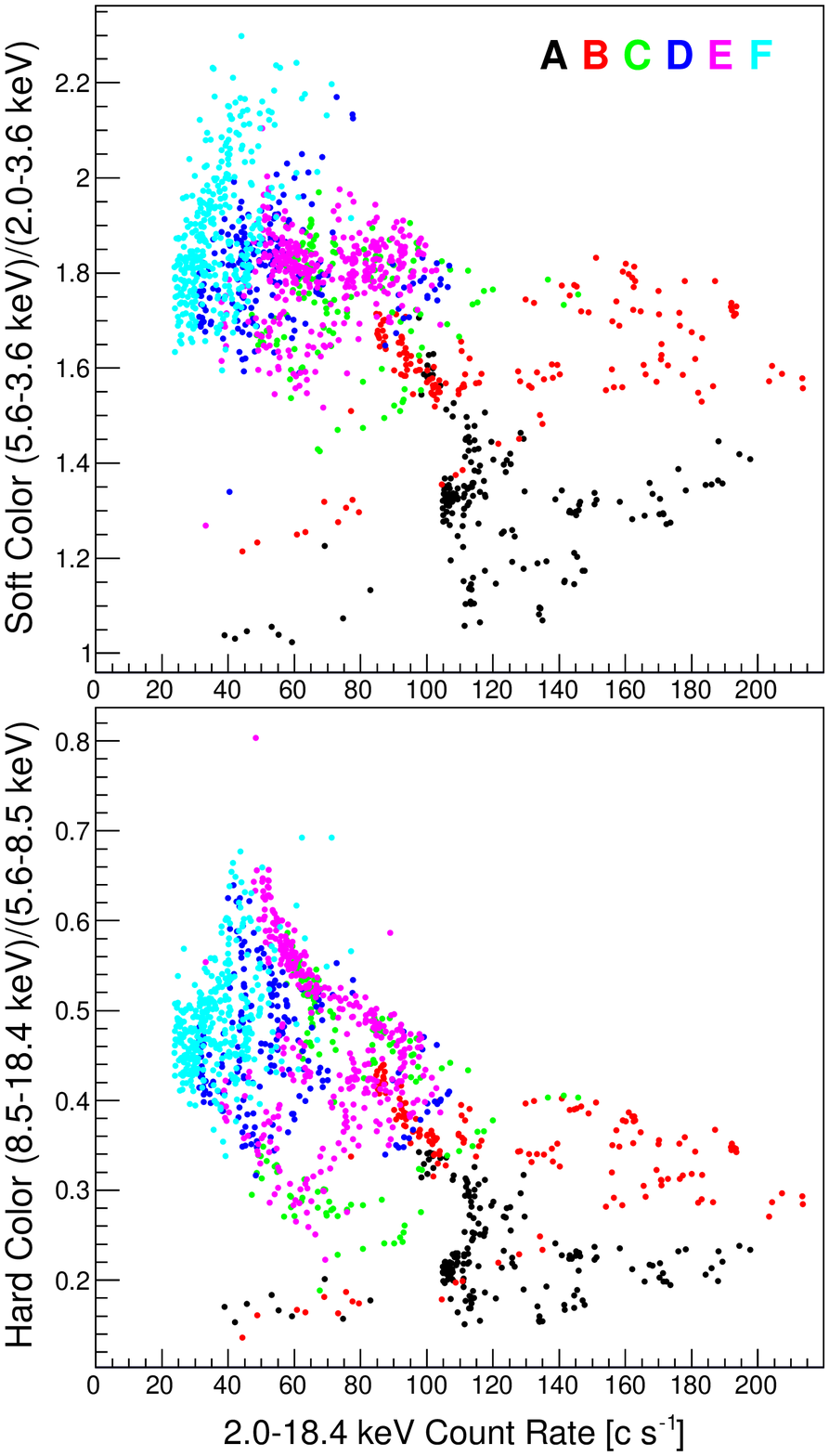}
\caption{ Soft-color versus intensity (top) and hard-color versus
  intensity (bottom) diagrams extracted from the RXTE-PCA data.  Data
  intervals of A, B, C, D, E, and F are colored similarly as
  in figure \ref{fig:lc_hr} and figure \ref{fig:ccdigaram}.  }
\label{fig:cnt_hr_relation}
\end{minipage}
\end{figure*}

To better grasp the spectral changes associated with the intensity
variations, we plot several CDs in figure \ref{fig:ccdigaram}, representing
the correlation between the two colors extracted
from the RXTE-PCA data.  There, data from the six intervals defined in
figure \ref{fig:lc_hr} are shown separately.
All the six CDs apparently exhibit Z-like shapes as often 
observed in bright NS
LMXBs (e.g. \cite{2006csxs.book...39V}).  This strongly supports the
source identification as a NS X-ray binary.
These CDs are also classified into two groups.  The Int-C behavior
resembles that of Int-E, while Int-D resembles Int-F.  This agrees
with the report by \citet{2011ATel.3650....1H} on this source,
that the CD track changed between 2011 February (MJD$\sim$55600) and
2011 September (MJD$\sim$55800).
According to two subgroups of the Z-like CDs, 
represented by Cyg X-2 and Sco X-1 
(e.g. \cite{2010ApJ...719..201H}), the CDs of Int-A, B, C, and E are
classified into the Cyg X-2 group, while those of Int-D and F are into
the Sco X-1 group.

In figure \ref{fig:cnt_hr_relation}, we plot the two colors against
the intensity, to create two HIDs. There, the six intervals are
specified with the same color scheme as in figure \ref{fig:ccdigaram}.
As \citet{2011ATel.3650....1H} reported, The change of the HID tracks
among the six intervals is also clearly seen.

\subsection{Color-Color and Hardness-Intensity Diagrams}
\label{sec:ana_cd_hid}

\subsection{Search for X-ray Burst Activity}

From multi-wavelength observations reported so far (section
\ref{sec:intro}) and the results of data analysis presented above,
MAXI J0556$-$332 is considered most likely to be a NS X-ray binary.
Transient NS X-ray binaries sometimes emit type-I X-ray bursts,
characterized by a fast rise in a few seconds and an exponential decay
in a few tens of seconds (e.g. \cite{1993SSRv...62..223L}).
Therefore, we searched the entire RXTE/PCA light curves for type-I
X-ray bursts using 1-s time bins.  However, we did not find any
burst-like event with a count-rate increase higher than 50 c s$^{-1}$
PCU$^{-1}$ in the 2--20 keV band.

Even if the source is located at the distance of 35 kpc, which is the
farthest limit estimated by \citet{2011ATel.3650....1H}, we would have
detected $\sim 98$ c s$^{-1}$ with the RXTE/PCA from a typical X-ray
burst, assuming a blackbody spectrum with a temperature of 2.1 keV and
a luminosity as high as the Eddington luminosity of $2\times 10^{38}$
erg s$^{-1}$ as seen in the typical type-I burst.  Such a burst should
have been observed with the RXTE/PCA.

\subsection{Initial Transition Behavior}

\begin{figure}
\begin{center}
\FigureFile(8.6cm,){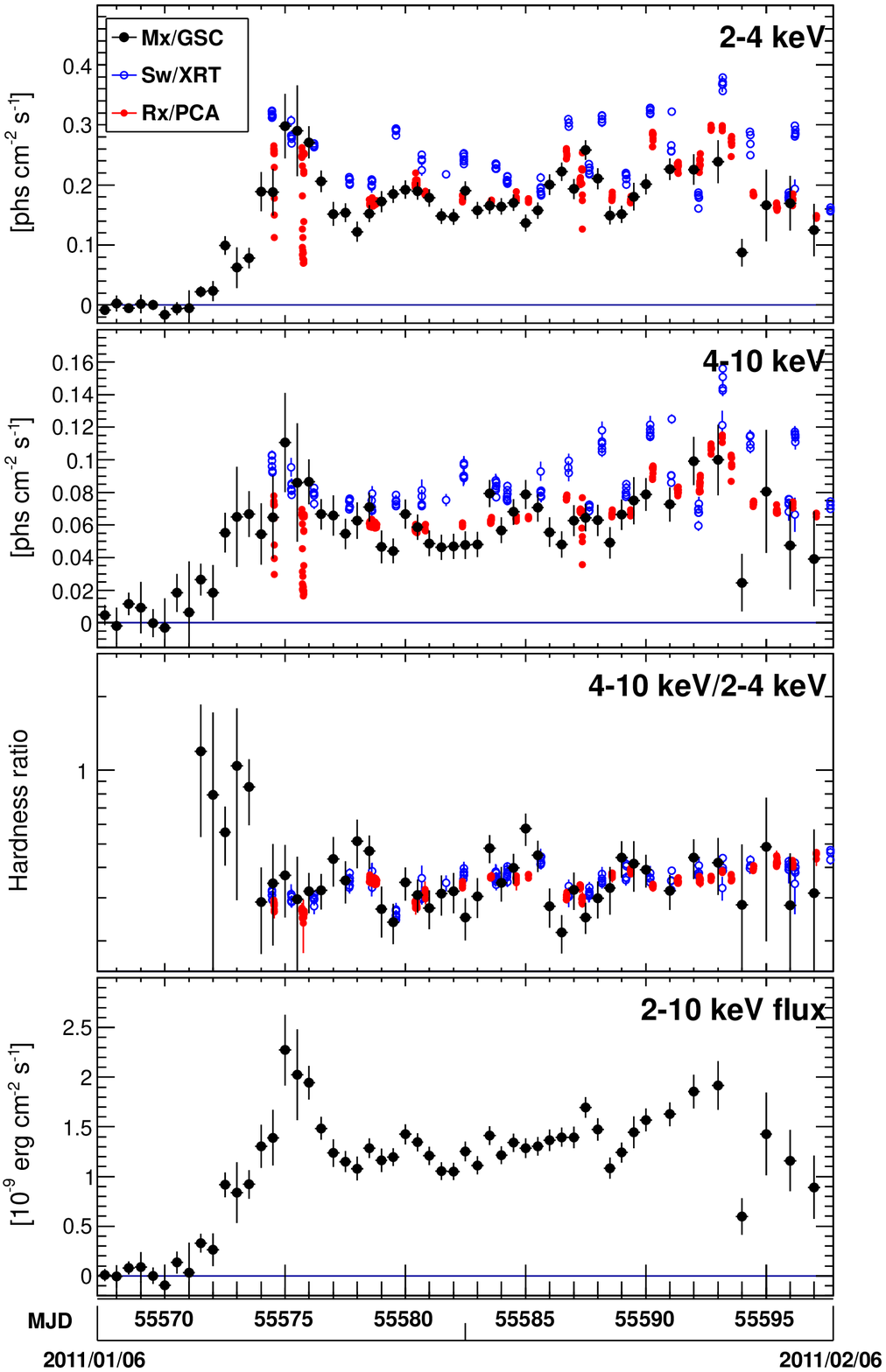}
\end{center}
\caption{ 

  Two-band MAXI/GSC light curves (black) during the initial phase,
  plotted since five days before the first detection.  Each data point
  is an average over 12 hours.  The Swift/XRT and RXTE/PCA data in
  256-s time bin are overlaid in blue and red, respectively.  Hardness
  ratio (4--10 keV / 2--4 keV) and the 2-10 keV flux are shown in the
  third and the fourth panels, respectively.

}
\label{fig:lc_hr_init}
\end{figure}

The initial emergent phase for about three days of MAXI J0556$-$332
was covered only by the MAXI/GSC all-sky survey.  To clarify the
source behavior meanwhile, we extracted 2--4 keV and 4--10 keV
lightcurves in a 12-h time bin, and present them in figure
\ref{fig:lc_hr_init}.  Like in figure \ref{fig:lc_hr}, the GSC count
rates were converted to the incident photon fluxes, to make the
results approximately free from the instrumental responses.
There, the Swift/XRT and the RXTE/PCA data are
superposed, together with the hardness ratio and the 2$-$10 keV flux.
The Swift/XRT and RXTE/PCA data points, each covering typically
0.5--2 ks (plotted with a bin width of 256 s), are significantly
more scattered than the 12-h averaged GSC data, presumably
due to short-term (e.g. $< 1$ h) variations.
However,
the hardness ratios are always consistent among the three missions.  
This implies that the spectral shape did not
change largely on a time scale of a day or shorter,
at least in the initial phase.

The MAXI/GSC hardness ratio clearly shows a step-like decrease at
MJD$=$55574.0 before the RXTE and Swift observations started.
Therefore, the source is considered to have made a hard-to-soft
transition at this epoch, when the flux was still rising rapidly.  The
flux then peaked at MJD$=$55575.5 in the presumable soft state.

\subsection{X-ray Spectral Evolution} 
\label{sec:ana_spec}

We carried out X-ray spectral analysis over the 0.5--30 keV broad
band, by combining the Swift/XRT and RXTE/PCA data which are averaged
over each observation (0.5--2 ks typically).
Although these Swift and RXTE observations were not exactly
simultaneous, any spectral change within a day is considered small as
already noticed in section \ref{sec:longlc}.
We thus selected pairs of Swift and RXTE observations carried out
within 12 hours, and performed their simultaneous fits.  To avoid
possible inconsistency between the two instruments, we left the PCA
versus XRT normalization, $f_{\rm PCA/XRT}$, free, and discarded those
pairs of which the value of $f_{\rm PCA/XRT}$ is deviated from the
average ($=1.12$) by more than 30\%.
The logs of the selected observations, totaling 72 pairs, are
summarized in table \ref{tab:spec_obslog}.  All spectral fitting was
carried out using Xspec version 12.7.0, released as a part of HEASOFT
6.11.

\begin{figure*}
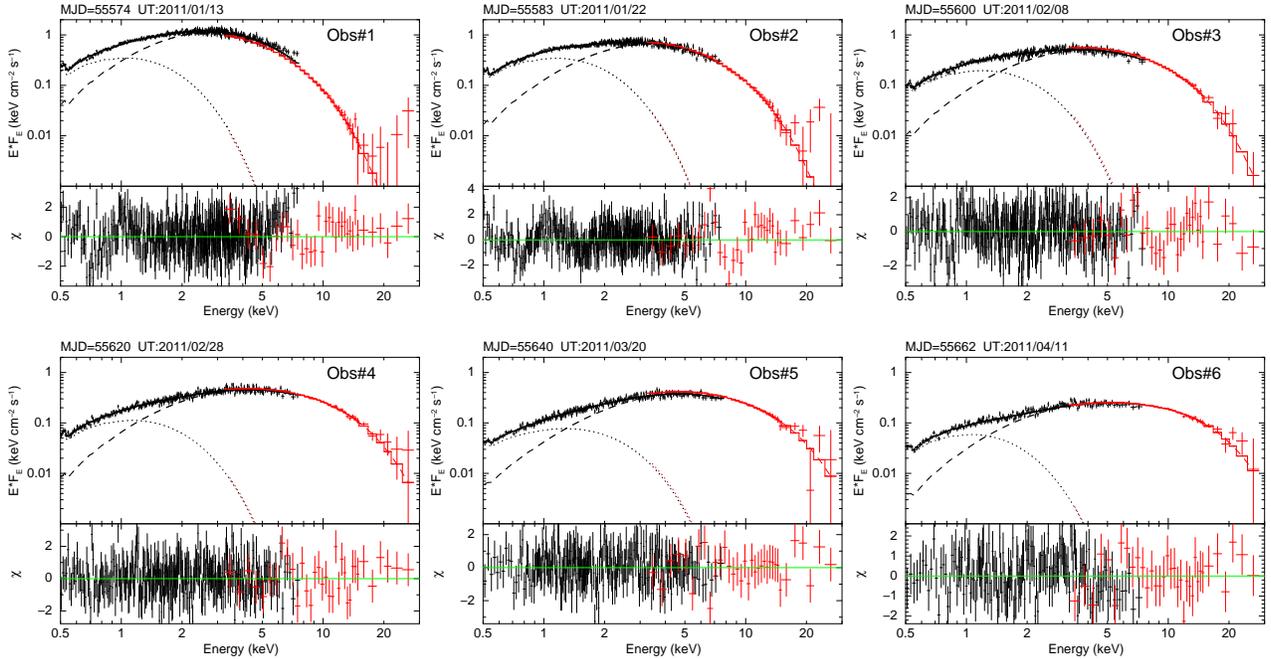

\begin{center}

\FigureFile(5.5cm,){fig6_0.eps}
\FigureFile(5.5cm,){fig6_1.eps}
\FigureFile(5.5cm,){fig6_2.eps}

\FigureFile(5.5cm,){fig6_3.eps}
\FigureFile(5.5cm,){fig6_4.eps}
\FigureFile(5.5cm,){fig6_5.eps}

\end{center}
\caption{ Examples of unfolded $\nu f\nu$ spectra of MAXI J0556$-$332
  obtained by the Swift/XRT (black) and the RXTE/PCA (red).  The
  best-fit {\tt wabs*(diskBB+nthcomp)} model and its residuals are
  shown together.  The model parameters are summarized in table
  \ref{tab:spec_parameters}.  
  Epochs of these six observations are indicated in 
  the top panel of figures
   \ref{fig:cutoff_param}a.  }
\label{fig:spectra}
\end{figure*}

\begin{figure*}
\begin{center}
\FigureFile(8.4cm,){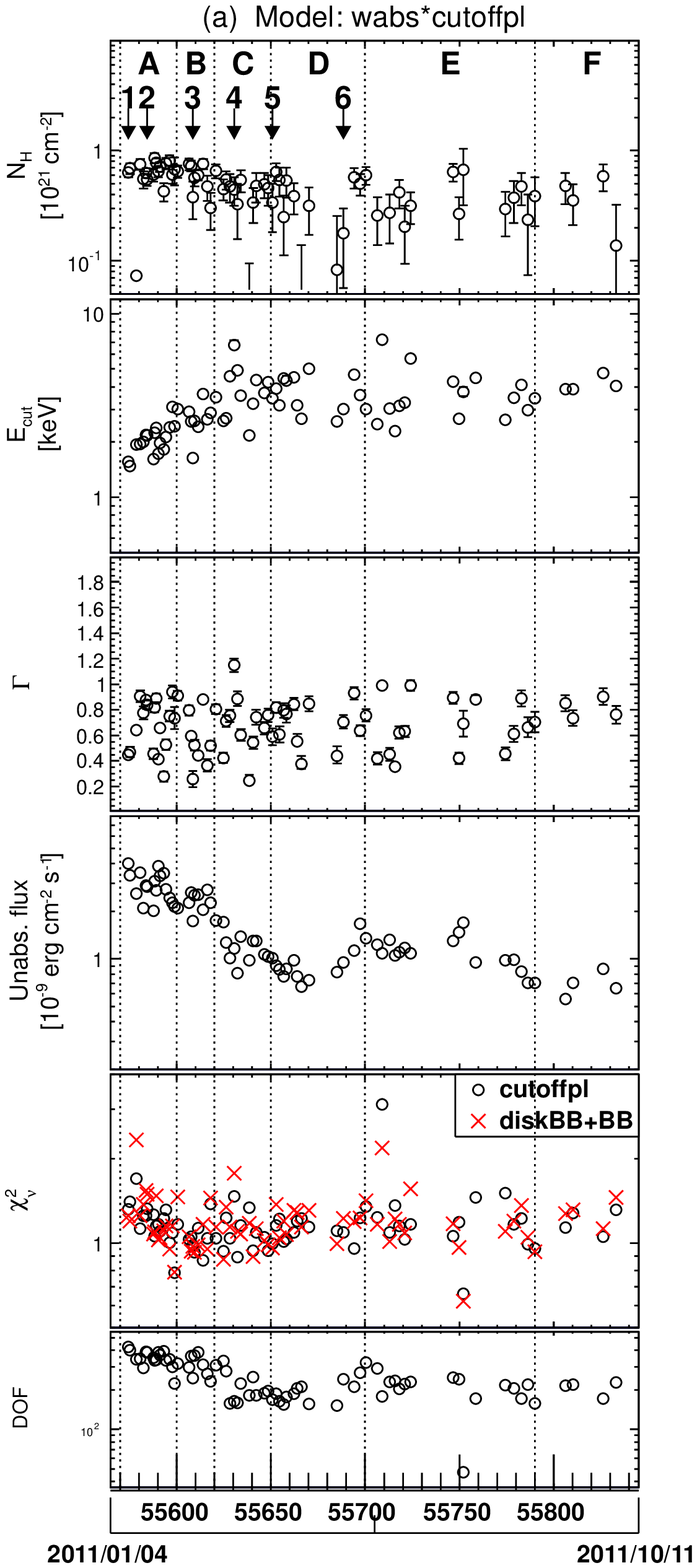}
\FigureFile(8.4cm,){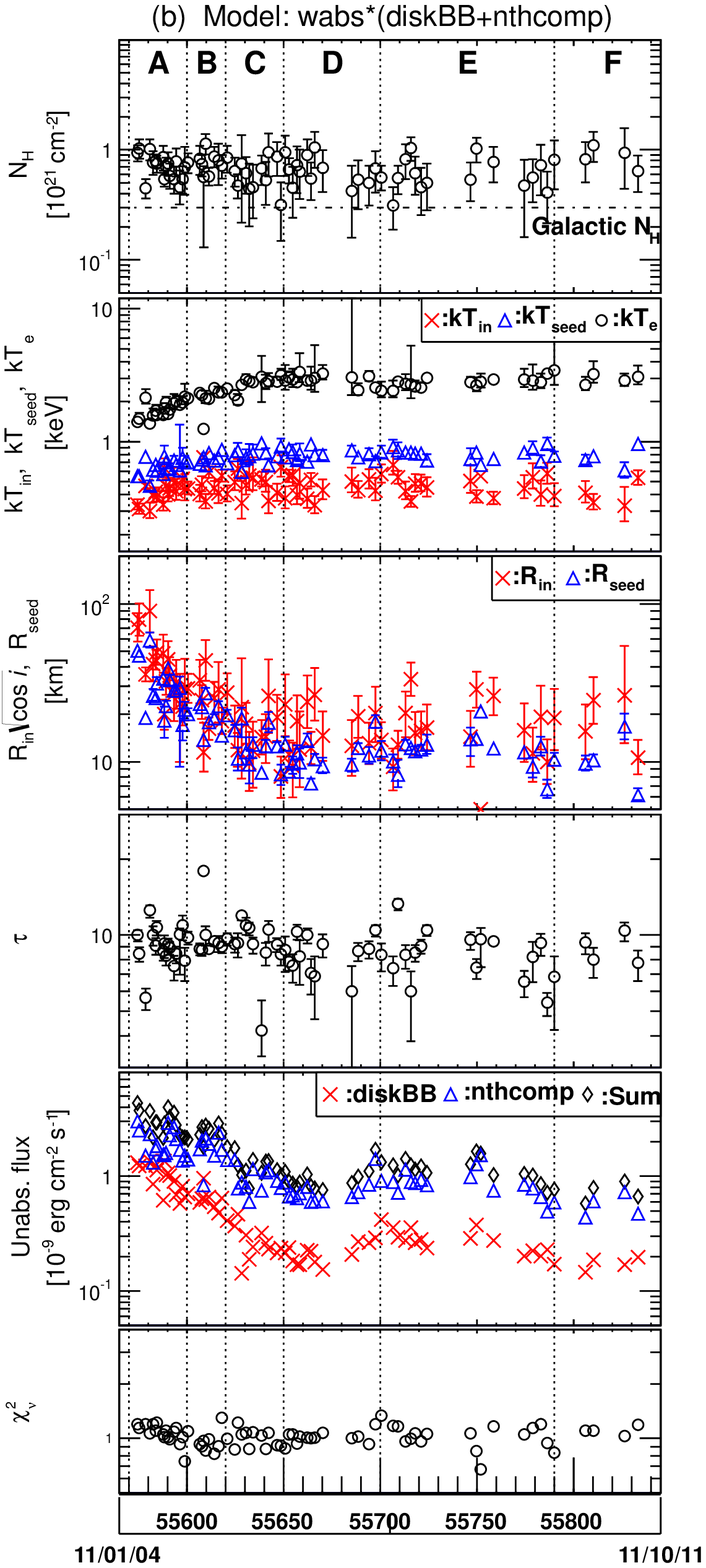}
\end{center}
\caption{ 
  Evolution of the best-fit spectral parameters, jointly
  determined with the Swift/XRT and the RXTE/PCA data.
  (a) The results from the empirical {\tt wabs*cutoffpl} fits.
  From top to bottom, the absorbing column density,
  the cutoff energy, the power-law index, 
  the absorption-corrected model flux,
  the $\chi^2_\nu$ values, and the degree of freedom.
  In the fifth panel, the fit goodness with an alternative
  {\tt wabs*(diskBB+BB)} is shown in (red) cross. 
  (b) The case with the more physical {\tt wabs*(diskBB+nthcomp)}
  fits. In the third panel, $R_{\rm seed}$ and $R_{\rm in}\sqrt{\cos i}$ are
  calculated in the assumed source distance of 10 kpc.
  
}
\label{fig:cutoff_param}
\label{fig:param_compttdisk}

\end{figure*}

Figure \ref{fig:spectra} shows the XRT and PCA spectra in their
unfolded $\nu f\nu$ form, from six typical observations, as indicated
with arrows in the top panel of figure \ref{fig:cutoff_param}a.
Including the six representatives, all the obtained spectra were found
to show a featureless continuum, without any significant emission or
absorption features.
We first attempted to fit them with a simple continuum model of either
a power law ({\tt PL}), a blackbody ({\tt BB}), a broken power-law
({\tt bknpower}), or a power law with a high-energy exponential cutoff
({\tt cutoffpl}), all multiplied with photoelectric
absorption ({\tt wabs}, \cite{1983ApJ...270..119M}) with free
column density $N_{\rm H}$.  
Then, 50 out of the 72 spectral pairs were reproduced by the {\tt
  wabs*cutoffpl} model with reduced chi-squared ($\chi^2_\nu$) within
the 99\% confidence limits, while none of the other models were as
successful on any data set.
Figure  \ref{fig:cutoff_param}a plots time evolution of best-fit {\tt
  wabs*cutoffpl} model parameters, together with absorption-corrected
0.1--100 keV fluxes.  Thus, the photon index $\Gamma$ was in the range
of 0.4--1 throughout, while the cutoff energy $E_{\rm cut}$ changed
from $\sim 1.5$ keV to $\sim 5$ keV.  The flux decreased from $4\times
10^{-9}$ to $5\times 10^{-10}$ erg cm$^{-2}$ s$^{-1}$ over the
10 months in agreement with the light curve (figure \ref{fig:lc_hr}). The
best-fit parameters for the spectra in figure \ref{fig:spectra} are
summarized in table \ref{tab:spec_parameters}.

The spectrum, represented by a cutoff power-law with $\Gamma=$ 0.4--1
and $E_{\rm cut}=$ 1.5--5 keV, roughly agrees with those of typical
bright NS-LMXBs. The cutoff energy could represent the temperature of
blackbody radiation from a neutron-star surface
(e.g. \cite{mitsuda1984PASJ}, \cite{1989PASJ...41...97M}).
We thus attempted a canonical two-component model for bright NS-LMXBs,
consisting of a multi-color-disk blackbody ({\tt diskBB} in Xspec
terminology) and a blackbody ({\tt BB}) \citep{mitsuda1984PASJ}.
However, as shown in figure  \ref{fig:cutoff_param}a (fifth panel) and
table \ref{tab:spec_parameters}, the best-fit $\chi^2_\nu$ values of
the {\tt wabs*(diskBB+BB)} fits were no better than those with the
{\tt wabs*cutoffpl}.
The residuals revealed excess features at both low and high energies,
which can be considered as a signature of a Comptonization process.
%
%
Following \citet{2007ApJ...667.1073L} and \citet{2009ApJ...696.1257L},
we hence tried to add a broken power-law function approximating the
Comptonized component to the canonical two-component model, but the
model {\tt wabs*(diskBB+BB+bknpower)} did not improve the fit
significantly.

\input{obslogtable.tex}

We then examined more rigorously the Comptonization emission process,
employing a thermally Comptonized continuum model, {\tt nthcomp} in
Xspec terminology (\cite{1996MNRAS.283..193Z};
\cite{1999MNRAS.309..561Z}). It describes Comptonized emission arising
when thermal seed photons with a {\tt BB} or a {\tt diskBB} spectrum
is up-scattered by a hot thermal electrons with a temperature
$kT_{e}$. This model uses an asymptotic power-law photon index
$\Gamma$, which is related with $kT_{e}$ and the scattering optical
depth $\tau$ as
\begin{equation}
\Gamma = \left[\frac{9}{4} + \frac{1}{\frac{kT_{\rm e}}{m_{\rm e}c^2}\tau(1+\frac{\tau}{3})}\right]^{\frac{1}{2}}-\frac{1}{2} 
\label{equ:comptau}
\end{equation}
\citep{1980A&A....86..121S}.

The {\tt nthcomp} model alone, with seed photons of neither {\tt BB}
nor {\tt diskBB}, was able to fit the observed spectra, even if the
$N_{\rm H}$ was left free.  The residuals showed a large discrepancy
in the soft X-ray band below 2 keV.  Adding another soft thermal
component, of which the spectral shape is the same as that of the seed
photons, did not improve the fit.  However, when the temperature of
the soft component was allowed to be free, and hence is different from
the seed-photon temparature, the Comptonization plus soft-component
model became acceptale.
Therefore, the seed photons for Comptonization and the additional soft
component must be of different origin.

When the additional soft component is represented by a {\tt BB} model,
the fit alwasys made the absorption column density $N_{\rm H} <
0.1\times 10^{21}$ cm$^{-2}$.  This upper limit is significantly lower
than the Galactic H$_{\rm I}$ density of $0.29\times 10^{21}$
cm$^{-2}$ in the source direction \citep{2005A&A...440..775K}.  
This would contradict the high Galactic latitude
($|b|=$\timeform{25D2}) and the suggested large distance, which place
MAXI J0556$-$332 well outside the Galactic disk.  On the other hand,
an alternative model employing a {\tt diskBB} for the soft component
always gave the best-fit $N_{\rm H}$ which is larger than the Galactic
H$_{\rm I}$.  Therefore, we hereafter use the {\tt diskBB} model for
the soft component.

Then, how about the seed-photon spectrum?
If the optical depth of the Comptonizing medium is high enough ($\tau
\gg 1$), whether the seed photon spectrum follows a {\tt BB} or a {\tt
  diskBB} model causes little difference in the emergent spectrum, and
therefore are difficult to distinguish.
Hence, we assume that the seed photons have a {\tt BB} spectrum.  In
section \ref{sec:discuss_specmodel}, We discuss the validity of this
model from the obtained best-fit parameters.

Using the {\tt wabs*(diskBB+nthcomp)} model with a {\tt BB}
seed-photon source, we fitted all the spectrum pairs, and obtained
successful fits as exemplified in figure \ref{fig:spectra}. Then,
assuming that the Comptonization process conserves the photon
number in the original {\tt BB} radiation, we estimated the radius of
the {\tt BB} seed-photon sphere $R_{\rm seed}$ according to the
formula of
\begin{eqnarray}
&& F_{\rm nthcomp} (T_{\rm seed}, R_{\rm seed}, d) ~~~~~ \left(\, {\rm photons \; cm^{-2} \; s^{-1}}\,\right) \nonumber \\
&& = \int f_{\rm BB}(E, T_{\rm seed}, R_{\rm seed}, d) dE \nonumber \\
&& = 20.1\left(\frac{kT_{\rm seed}}{1\; \rm keV}\right)^{3} \left(\frac{R_{\rm seed}}{1\; \rm km}\right)^2 \left(\frac{d}{10\; \rm kpc}\right)^{-2},
\label{equ:rseed}
\end{eqnarray}
where $F_{\rm nthcomp}(T_{\rm seed}, R_{\rm seed}, d)$
is the incident photon flux calculated
from the best-fit model parameters in Xspec,
$T_{\rm seed}$ is the seed-photon {\tt BB} temperature, 
$d$ is the source distance, 
and $f_{\rm BB}(E, T_{\rm seed}, R_{\rm seed},
d)$ is the photon-flux spectrum of a {\tt BB} emission.
%
%
Including the values of $R_{\rm seed}$, the best-fit
model parameters from the six representative spectra are given in
table \ref{tab:spec_parameters}.

Figures \ref{fig:param_compttdisk}b show temporal variations of the
best-fit parameters, including $\tau$ and $R_{\rm seed}$ derived using
equations (\ref{equ:comptau}) and (\ref{equ:rseed}) respectively, and
the unabsorbed fluxes of the {\tt diskBB} and {\tt nthcomp}
components.  The values of $R_{\rm seed}$ and the {\tt diskBB}-model
parameter, $R_{\rm in}\sqrt{\cos i}$, related to the disk inner radius $R_{\rm
  in}$ and the inclination $i$, are calculated 
assuming $d=10$ kpc. The actual radii are proportional to the source
distance.

\input{spectable.tex}


\section{Discussion}
\label{sec:discussion}

MAXI J0556$-$332 is a new X-ray transient which appeared on 2011
January 11.  
Using the MAXI/GSC, Swift/XRT, and RXTE/PCA data, we studied the
intensity and spectal evolution of this source over the entire active
period for more than one year.

\subsection{Source Identification from X-ray Spectrum}
\label{sec:discuss_identity}

The wide-band (0.5--30 keV) X-ray spectra obtained by the Swift/XRT
and XTE/PCA are featureless, and can be
approximated by a cutoff power-law with $\Gamma\approx 0.4-1$ and
$E_{\rm cut}\approx 1.5-5$ keV.  They are better fitted with
a two-component model that consists of a multi-color-disk
blackbody and a thermally Comptonized blackbody.  
This spectral model has often been used to describe 
X-ray emission from LMXBs
containing a weakly-magnetized NS (e.g. \cite{2002MNRAS.337.1373G};
\cite{2012arXiv1201.5891S}) or a BH candidate
(e.g. \cite{2004MNRAS.349..393D}). 

NS LMXBs have been largely classified into ``Z-type'' and ''atoll-type''
sources according to their CD-track shapes \citep{1989A&A...225...79H,
  2006csxs.book...39V}. The former objects are persistently as bright as
$\ledd$,
and sometimes show rapid flux variations up to a factor of $\sim
5$. The latter are fainter than the former, and often appear
as transients or reccurents.  The luminosity of atoll sources
varies over a large range from $\sim 10^{-3}$ to $\sim 0.2$ times $\ledd$,
accompanied by spectral state changes.  
The higher-energy cutoff of 1.5--5 keV
of MAXI J0556$-$332, represented by $E_{\rm cut}$ in
the {\tt wabs*cutoffpl} model or $kT_{\rm e}$ in the {\tt
  wabs*(diskBB+nthcomp)} model, agrees well with those of the Z
sources such as Cyg X-2 \citep{2002A&A...386..535D}, Sco X-1
\citep{2007ApJ...667..411D}, GX 5$-$1 \citep{2011ApJ...743L..31S}, GX
17$+$2 \citep{2005A&A...434...25F}, and GX 349$+$2
\citep{2001ApJ...554...49D}, whose luminosities are always
$>0.5\,\ledd$. The $E_{\rm cut}$ values of atoll sources are also
below 5 keV in the soft state with the luminosity higher than $0.1\,\ledd$, 
while they tend to increase up to $>$ 15
keV in the hard state with the luminosity of the order of
$0.01\,\ledd$; the latter expamples include Aql X-1
\citep{2007ApJ...667.1073L,2012arXiv1201.5891S}, 4U 0614+09
\citep{1994ApJ...431..826S}, 4U 1608$-$52
\citep{2002MNRAS.337.1373G,2007ApJ...667.1073L,2011ApJ...738...62T},
4U 1705-44 \citep{2002ApJ...576..391B,2010ApJ...719.1350L}, and 4U
1728$-$34 \citep{2011MNRAS.416..873T}.
Therefore, if the source is a NS LMXB, the X-ray luminosity should be
higher than $0.1\,\ledd = 1.8\times 10^{37}$ erg s$^{-1}$ in the
observed period.  The absorption-corrected model flux, which changed
from $4\times 10^{-9}$ to $5\times 10^{-10}$ erg cm$^{-2}$ s$^{-1}$,
constrains the source distance to be $d>17$ kpc, in agreement with
the estimate of $d=$20--35 kpc by \citet{2011ATel.3650....1H}
deduced from the luminisity at a CD-track transition.

Galactic BH X-ray binaries have been observed mostly in either of the
two major spectral states, the low/hard or the high/soft state. These
spectra are represented by a combination of thermal emission from the
accretion disk and a Comptonized harder component (although detailed
model parameters differ significantly between the tow states).  The
obtained $E_{\rm cut}$ of 1.5--5 keV seems to be too high as the
temperature for the accretion disk in BH binaries, which is typically
below 1 keV, and is also too low for the cutoff energy of the
Comptonized component, which is usually higher than 50 keV
(e.g. \cite{2006csxs.book..157M}).  However, the peculiar Galactic BH
binary, GRS 1915$+$105, sometimes shows in the bright phases
such a spectrum as represented
by {\tt diskBB} with $kT_{\rm in}\sim$2 keV 
\citep{2004MNRAS.349..393D}.  Therefore, the possibility of a BH X-ray
binary cannot be completely ruled out from the spectral parameters 
alone. 
As has been seen in the CDs and HIDs (section \ref{sec:ana_cd_hid}),
the spectral time evolution, to be explored below, gives us another
helpful information to identify the source nature.

\subsection{Emission Geometry in NS-LMXB Scenario}
\label{sec:discuss_specmodel}

According to the widely accepted picture of NS-LMXBs
(e.g. \cite{mitsuda1984PASJ}, \cite{1989PASJ...41...97M}), the two
continuum components in the {\tt wabs*(diskBB+nthcomp)} model are
interpreted as a thermal emission from the accretion disk ({\tt
  diskBB}), and a blackbody emission from the NS surface
(the boundary layer) modified through Compotonization 
by surrounding hot electrons ({\tt nthcomp}).
In figure \ref{fig:param_compttdisk}b, time evolution of the parameters
of these physical components are plotted.

In order for the {\tt diskBB} model to be physical, 
its inner disk radius has to be larger
than the NS radius, $\sim 10$ km.
We may derive a realistic estimate of the inner disk radius, $r_{\rm in}$,
from the model parameter $R_{\rm in}\sqrt{\cos i}$ as
\begin{eqnarray}
r_{\rm in} &=& \xi \kappa^2 R_{\rm in} \nonumber \\
&=& 1.2 \left( \frac{\xi}{0.41} \right) \left( \frac{\kappa}{1.7} \right)^2 \left(\frac{d_{10}}{\sqrt{\cos i}} \right) \cdot R_{\rm in}\sqrt{\cos i},
\end{eqnarray}
where
$\xi=0.41$ is a correction factor for the inner boundary condition 
\citep{1998PASJ...50..667K, 2000ApJ...535..632M}, $\kappa=1.7$ is the
standard color hardening factor \citep{1995ApJ...445..780S}, 
and $d_{10}$ is the source distance in units of 10 kpc.
Here we adopted the same canonical $\xi$ and $\kappa$ values as for 
the accretion disk around a black hole.
We find that $r_{\rm in}$
changed from $\sim 80\, (d_{\rm 10}/\sqrt{\cos i})$
km to $\sim 20\, (d_{\rm 10}/\sqrt{\cos i})$ km
across the observations.

If the source distance is $>17$ kpc as estimated in section
\ref{sec:discuss_identity}, the condition of $r_{\rm in}>10$ km is
always satisfied.  The best-fit parameters also satisfy the expected
relations of $R_{\rm seed} < r_{\rm in}$ and $kT_{\rm seed} > kT_{\rm
  in}$.  Thus, the derived model parameters are reasonable, although
the range of $kT_{\rm in}\sim 0.3-0.6$ keV seems to be slightly lower
than the typical values as obtained in Cyg X-2
\citep{2002A&A...386..535D}, Aql X-1 \citep{2012arXiv1201.5891S}, and
4U 1705-44 \citep{2010ApJ...719.1350L} from the BeppoSAX or Suzaku
X-ray spectra covering the 0.2--50 keV band.

The seed photons of the Comptonized emission in the {\tt nthcomp}
component are considered to mainly originate from a region near 
the NS surface, to be called the boundary layer.
The obtained temperatures of $kT_{\rm seed}\approx 0.7$ keV and
$kT_{\rm e}\approx 1.5-3$ keV, agree with those of NS LMXBs 
(either Z sources or atoll sources) in the soft state,
derived using the same
spectral model (e.g. \cite{2002MNRAS.337.1373G};
\cite{2005A&A...434...25F}; \cite{2011MNRAS.416..637R};
\cite{2012arXiv1201.5891S}).
As seen in figure \ref{fig:param_compttdisk}b, however, 
the estimated radius of the {\tt BB} seed photons sphere, $R_{\rm
  seed}$,
changed from $\sim 40\, d_{\rm 10}$ km to $\sim 10\, d_{\rm 10}$ km.
If the source distance is $>17$ kpc as estimated in section
\ref{sec:discuss_identity}, 
$R_{\rm seed}$ becomes significantly larger than the typical NS
radius of $\sim 10$ km. This could be explained if the emission from the
accretion disk also contributed to the seed photons.

As shown so far, the combination of a disk emission and a Comptonized
blackbody component can consistently explain the spectral parameters
of MAXI J0556$-$332.
Figure
\ref{fig:ns_schemaric_view} illustrates a possible source geometry
implied by the model.
The relatively low values of $kT_{\rm in}$ may be explained by the
suggested obscuration of the innermost disk region by the Compton
corona.

\subsection{Long-Term Spectral Evolution}

\begin{figure}
\begin{center}
\FigureFile(8.6cm,){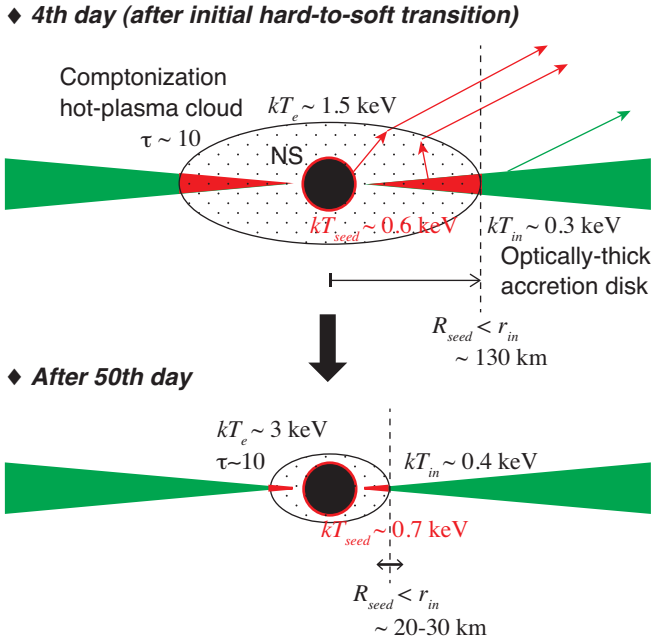}
\end{center}
\caption{ A schematic view of X-ray emission region including the NS
  surface, the accretion disk, and the surrounding hot-plasma cloud,
  implied by the best-fit spectral model on the fourth day (top panel)
  and after 50th day (bottom panel) from the activity onset.
  A source distance of 26 kpc is employed to calculate $R_{\rm seed}$
  and $r_{\rm in}$.  }
\label{fig:ns_schemaric_view}
\end{figure}

Over the initial $\sim 70$ days, figure \ref{fig:param_compttdisk}b
reveals a decrease in both $r_{\rm in}$ and $R_{\rm seed}$, and increase
in $kT_{\rm e}$.
These changes are considered to reflect the evolution in the X-ray
emission region 
as illustrated in figure \ref{fig:ns_schemaric_view}.

\citet{2003A&A...410..217G} and \citet{2006A&A...453..253R} reported
that the spectrum of the boundary layer at the NS surface can be
approximated by a saturated ($\tau\gg 1$) Comptonization model with
$kT_{\rm e}\approx 2-4$ keV, to which our results agree.
\citet{2001ApJ...547..355P} numerically explored the same issue, 
and suggested that the boundary laryer would expand at luminosities near
$\ledd$. In such a condition, the optical depth of the Comptonization
media would be further increased.  The observed relations between the
flux and spectrum agree with the expected tendency, although more
detailed quantitative study is out of the scope of this paper.

\subsection{State Transition in Color-Color Diagram}
\label{sec:disc_colcol}

State transitions are considered to reflect changes of the physical
condition in the emission region including the NS surface (or the
boundary layer) and the accretion disk, mainly in response to changes
in the luminosity.
On the basis of the Z source features seen in the CD/HID tracks
extracted from the RXTE/PCA data, \citet{2011ATel.3650....1H} reported
that MAXI J0556$-$322 is the third transient Z source after XTE
J1701$-$462 \citep{2010ApJ...719..201H} and IGR J17480$-$2446
\citep{2010ATel.2952....1A}.
They also reported on a source transition between the two types of Z
source tracks, from Cyg-like to Sco-like substates. This kind of
transition has been observed previously only from XTE J1701$-$462,
which exhibited all three kinds of the CD/HID tracks (Cyg-like Z,
Sco-like Z, and atoll) during the decay phase of its outburst in
2006--2007 \citep{2010ApJ...719..201H}.
Actually, in figure \ref{fig:ccdigaram} and figure
\ref{fig:cnt_hr_relation}, we confirmed that these tracks are
classified into either Cyg-like Z (int-A, B, C, and E) or Sco-like Z
(int-D and F). These CDs and HIDs resemble those of XTE J1701$-$462 in
the 2006--2007 outburst \citep{2007ApJ...656..420H}.

The other two transient Z sources, XTE J1701$-$462 and IGR
J17480$-$2446, showed type-I X-ray bursts (\cite{2009ApJ...699...60L};
\cite{2011ApJ...730L..23C}).
However, we did not find any burst-like events.

\subsection{Distance Estimate and Source Location}

\citet{2011ATel.3650....1H} estimated the distance of MAXI J0556$-$322
to be 20--35 kpc assuming that the transition between the Cyg-like
and the Sco-like substates occurred at the same luminosity as that in XTE
J1701$-$462, whose distance is estimated to be 8.8 kpc from the
luminosity of its type-I X-ray bursts \citep{2009ApJ...699...60L}.  
In the energy-sorted light curves in figure \ref{fig:lc_hr}, 
the substate transition 
occurred when the 2.0--3.6 keV PCA count rate crossed $\sim$
14 c s$^{-1}$.  
For comparison, XTE J1701$-$462 exhibited the transition at
the PCA count rate of $\sim$ 80 c s$^{-1}$ in 2.0-2.9 keV,
or 121 c s$^{-1}$ in the 2.0--3.6 keV.
Assuming that the two sources make the transition at the same
intrinsic luminosity, MAXI J0556$-$322 is then estimated to be
2.9 times farther, located at $d\sim$ 26 kpc;
this reconfirms the estimate by
\citet{2011ATel.3650....1H}.
This estimate and the Galactic coordinates of
$(l,b)=(238^\circ.9, -25^\circ.2)$ indicate its location of 11 kpc
below the Galactic plane and 31 kpc away from the Galactic center.
This places the source on outskirts of the Galactic halo, where
the population of LMXBs is rather small \citep{2002A&A...391..923G}.

Assuming $d=26$ kpc, the observed maximum flux
of $4.3 \times 10^{-9}$ erg cm$^{-2}$ s$^{-1}$ in the initial phase
can be converted to the absorption-corrected luminosity of $3.5 \times
10^{38}$ erg s$^{-1}$.  This is twice as high as the Eddington limit
for the typical $1.4\,M_\odot$ NS, and agrees with the maximum
luminosity in some bright NS LMXBs such as Sco X-1
\citep{2007ApJ...667..411D}.

\subsection{Initial Transition}

The MAXI/GSC light curve in figure \ref{fig:lc_hr_init} reveals a
hard-to-soft state transition which occurred on the fourth day from
the brightening onset, at the middle of the initial brightening phase.
As seen in other transient NS LMXBs
\citep{asai_lmxb} as well as BH LMXBs \citep{2006MNRAS.370..837G},
this transition may be interpreted as the initial formation of an 
optically-thick accretion disk.
The observed X-ray flux just before the transition was $1.0 \times
10^{-9}$ erg cm$^{-2}$ s$^{-1}$ in the 2--10 keV band.

By comparing the Swift/BAT and the MAXI/GSC data, \citet{asai_lmxb}
investigated outburst rising behavior in the two transient NS
binaries, Aql X-1 and 4U 1608--52,
and revealed that the initial hard-to-soft state transition of these
sources occur at a 2--15 keV luminosity of $0.5\times 10^{36} -
2\times 10^{37}$ erg s$^{-1}$.
If the distance to MAXI J0556$-$322 is $>17$ kpc as estimated above,
the initial hard-to-soft transition is inferred to have occurred at a
luminosity of $>3.5\times 10^{37}$ erg s$^{-1}$. It is significantly
higher than that in Aql X-1 and 4U 1608--52.
Further discussion on this point requires more detailed consideration
of the inclination angle, and possible differences in magnetic
field strengths.

\section{Conclusion}

Throughout the active period which lasted for more than one year, the
0.5--30 keV X-ray spectra of the new X-ray transient, MAXI
J0556$-$322, obtained by the Swift/XRT and RXTE/PCA were successfully
represented by a two-component model, consisting of an optically-thick
thermal emission from an accretion disk with an inner-disk temperature
of $kT\approx 0.3-0.4$ keV, and a Comptonized emission of thermal seed
photons of $kT_{\rm seed}\approx 0.6-0.8$ keV by hot electrons with
$kT_{\rm e}\approx 1.5-3$ keV.  The obtained model parameters are
consistent with those of NS LMXBs when their luminosity is higher than
$\sim 0.1\,\ledd$.  
This, together with the source behavior on CDs,
constrains the source distance to be $>17$ kpc, most likely 26 kpc,
and place the source on outskirts of the Galactic halo.
The long-term spectral variations can be understood by evolution
of the accretion disk and the emission region on the NS.
The MAXI/GSC light curve revealed a hard-to-soft transition in
the middle of the initial
brightening phase. The transition luminosity would be 
significantly higher than those observed in other typical transient NS LMXBs.

\bigskip

We appreciate all MAXI science and operation team members for their
dedicated works to enable the science analysis of MAXI data.  We also
thank Drs. K. Makishima, K. Asai, and H. Takahashi for their
stimulating discussions.  This research has made use of RXTE data
obtained from the High Energy Astrophysics Science Archive Research
Center (HEASARC) provided by NASA/GSFC; and Swift data supplied by the
UK Swift Science Data Centre at the University of Leicester.  This
reseach work is partially supported by the Ministry of Education,
Culture, Sports, Science and Technology (MEXT), Grant-in-Aid for
Science Research 20244015.

\input{thebib.tex}

\end{document}

%% file: obslogtable.tex
\begin{table*}
\small
\renewcommand{\arraystretch}{0.6}

\caption{List of Swift/XRT and RXTE/PCA observation pairs used in the simultaneous spectal fits}
\label{tab:spec_obslog}
\begin{center}
\scriptsize 
\begin{tabular}{l|ccccc|ccccc}
\hline
\hline
 MJD & \multicolumn{5}{c}{Swift/XRT} & \multicolumn{5}{|c}{RXTE/PCA}                      \\
     & Obs ID    & Date$^\dagger$ & Start & End & Exp.[s] & Obs ID  & Date$^\dagger$ & Start & End & Exp.[s]  \\
\hline
 55574*  & 00031914001 & 01/13 & 10:51 & 11:14 & 1413 & 96371-01-02-00 & 01/13 & 12:32 & 13:22 & 2912\\
 55575   & 00031914002 & 01/14 & 06:10 & 06:31 & 1277 & 96371-01-01-00 & 01/14 & 16:44 & 19:06 & 5936\\
 55578   & 00031914005 & 01/17 & 14:23 & 14:42 & 1125 & 96371-01-03-00 & 01/17 & 12:13 & 19:24 & 15872\\
 55580   & 00031914007 & 01/19 & 16:09 & 16:24 & 891 & 96371-01-04-01 & 01/19 & 19:35 & 20:00 & 864\\
 55582   & 00031914009 & 01/21 & 10:10 & 10:33 & 1375 & 96371-01-05-00 & 01/21 & 08:52 & 09:17 & 1456\\
 55583*  & 00031914010 & 01/22 & 18:15 & 18:39 & 1466 & 96371-01-05-01 & 01/22 & 13:02 & 13:51 & 2800\\
 55584   & 00031914011 & 01/23 & 05:30 & 05:54 & 1480 & 96371-01-05-02 & 01/23 & 14:27 & 14:57 & 1616\\
 55587   & 00031914014 & 01/26 & 15:13 & 15:37 & 1442 & 96371-01-05-05 & 01/26 & 08:06 & 08:48 & 2352\\
 55588   & 00031914015 & 01/27 & 04:03 & 04:19 & 955 & 96371-01-05-06 & 01/27 & 13:56 & 14:36 & 2224\\
 55589   & 00031914016 & 01/28 & 04:08 & 04:27 & 1167 & 96414-01-01-00 & 01/28 & 08:29 & 09:25 & 3200\\
 55590   & 00031914017 & 01/29 & 04:13 & 04:28 & 938 & 96414-01-01-01 & 01/29 & 06:33 & 07:22 & 2752\\
 55591   & 00031914018 & 01/30 & 01:21 & 01:38 & 1006 & 96414-01-01-02 & 01/30 & 07:30 & 08:27 & 3200\\
 55593   & 00031914020 & 02/01 & 04:28 & 04:44 & 981 & 96414-01-01-07 & 02/01 & 04:05 & 04:21 & 800\\
 55594   & 00031914021 & 02/02 & 08:01 & 08:16 & 900 & 96414-01-01-05 & 02/02 & 11:09 & 11:43 & 1872\\
 55596   & 00031914023 & 02/04 & 04:48 & 05:07 & 1180 & 96414-01-02-00 & 02/04 & 02:33 & 02:54 & 1152\\
 55597   & 00031914024 & 02/05 & 16:04 & 16:17 & 760 & 96414-01-02-07 & 02/05 & 23:26 & 23:39 & 768\\
 55598   & 00031914025 & 02/06 & 19:30 & 19:34 & 252 & 96414-01-02-09 & 02/06 & 21:57 & 22:21 & 416\\
 55600*  & 00031914027 & 02/08 & 10:10 & 10:30 & 1165 & 96414-01-02-04 & 02/08 & 11:10 & 11:59 & 2368\\
 55606   & 00031914033 & 02/14 & 13:55 & 14:09 & 873 & 96414-01-03-03 & 02/14 & 13:03 & 15:00 & 2848\\
 55607   & 00031914034 & 02/15 & 13:49 & 14:08 & 1147 & 96414-01-03-04 & 02/15 & 10:54 & 12:53 & 3408\\
 55608   & 00031914035 & 02/16 & 13:52 & 14:08 & 959 & 96414-01-03-05 & 02/16 & 10:23 & 11:24 & 2896\\
 55609   & 00031914036 & 02/17 & 12:32 & 12:52 & 1208 & 96414-01-03-06 & 02/17 & 16:23 & 16:43 & 336\\
 55611   & 00031914038 & 02/19 & 07:43 & 08:05 & 1275 & 96414-01-04-01 & 02/19 & 05:48 & 09:52 & 8912\\
 55614   & 00031914039 & 02/22 & 01:35 & 01:54 & 1144 & 96414-01-04-04 & 02/22 & 07:34 & 10:51 & 6144\\
 55616   & 00031914040 & 02/24 & 04:58 & 05:18 & 1243 & 96414-01-04-06 & 02/24 & 11:19 & 11:40 & 880\\
 55618   & 00031914041 & 02/26 & 02:06 & 02:13 & 441 & 96414-01-05-01 & 02/26 & 11:57 & 13:26 & 3040\\
 55620*  & 00031914042 & 02/28 & 18:08 & 18:29 & 1260 & 96414-01-05-03 & 02/28 & 11:25 & 11:59 & 1872\\
 55624   & 00031914044 & 03/04 & 17:07 & 17:30 & 1373 & 96414-01-06-00 & 03/04 & 12:14 & 12:48 & 2000\\
 55626   & 00031914045 & 03/06 & 04:18 & 04:40 & 1309 & 96414-01-06-02 & 03/06 & 05:06 & 05:37 & 1872\\
 55628   & 00031914046 & 03/08 & 02:46 & 03:09 & 1398 & 96414-01-06-04 & 03/08 & 13:25 & 14:22 & 3184\\
 55630   & 00031914047 & 03/10 & 09:43 & 09:59 & 994 & 96414-01-06-06 & 03/10 & 09:28 & 09:49 & 1232\\
 55632   & 00031914048 & 03/12 & 04:40 & 04:49 & 538 & 96414-01-07-01 & 03/12 & 11:31 & 12:21 & 2944\\
 55634   & 00031914049 & 03/14 & 01:42 & 02:01 & 1148 & 96414-01-07-02 & 03/14 & 02:41 & 03:13 & 1888\\
 55638   & 00031914051 & 03/18 & 13:21 & 13:40 & 1165 & 96414-01-08-00 & 03/18 & 12:02 & 12:35 & 1952\\
 55640*  & 00031914052 & 03/20 & 15:12 & 15:32 & 1216 & 96414-01-08-02 & 03/20 & 10:59 & 11:20 & 1216\\
 55642   & 00031914053 & 03/22 & 02:30 & 02:46 & 968 & 96414-01-08-04 & 03/22 & 01:09 & 01:19 & 608\\
 55646   & 00031914055 & 03/26 & 10:45 & 11:04 & 1145 & 96414-01-09-00 & 03/26 & 07:55 & 08:47 & 2416\\
 55648   & 00031914056 & 03/28 & 12:30 & 12:46 & 972 & 96414-01-09-03 & 03/28 & 08:36 & 09:19 & 2128\\
 55650   & 00031914057 & 03/30 & 17:37 & 17:53 & 978 & 96414-01-09-05 & 03/30 & 10:52 & 11:10 & 592\\
 55652   & 00031914058 & 04/01 & 19:13 & 19:34 & 1277 & 96414-01-10-01 & 04/02 & 06:04 & 08:33 & 5216\\
 55654   & 00031914059 & 04/03 & 14:39 & 14:57 & 1092 & 96414-01-10-02 & 04/03 & 07:11 & 08:05 & 2592\\
 55656   & 00031914060 & 04/05 & 21:14 & 21:33 & 1169 & 96414-01-10-05 & 04/06 & 02:53 & 03:26 & 2016\\
 55658   & 00031914061 & 04/07 & 03:44 & 04:03 & 1156 & 96414-01-10-06 & 04/07 & 05:08 & 05:23 & 464\\
 55662*  & 00031914063 & 04/11 & 05:50 & 06:11 & 1258 & 96414-01-11-02 & 04/11 & 14:21 & 15:14 & 2272\\
 55664   & 00031914064 & 04/13 & 01:06 & 01:25 & 1143 & 96414-01-11-04 & 04/13 & 00:46 & 01:26 & 2320\\
 55666   & 00031914065 & 04/15 & 01:19 & 01:42 & 1392 & 96414-01-12-00 & 04/15 & 12:21 & 12:43 & 1072\\
 55670   & 00031914067 & 04/19 & 03:25 & 03:47 & 1340 & 96414-01-12-02 & 04/19 & 02:50 & 03:15 & 1488\\
 55684   & 00031914071 & 05/04 & 02:48 & 03:03 & 874 & 96414-01-14-02 & 05/03 & 20:49 & 21:11 & 1328\\
 55688   & 00031914072 & 05/07 & 11:16 & 11:33 & 1052 & 96414-01-15-01 & 05/07 & 07:51 & 08:48 & 3168\\
 55694   & 00031914074 & 05/13 & 03:29 & 03:48 & 1183 & 96414-01-16-00 & 05/13 & 06:42 & 07:25 & 1904\\
 55697   & 00031914075 & 05/16 & 10:20 & 10:39 & 1110 & 96414-01-16-03 & 05/16 & 06:50 & 07:24 & 1696\\
 55700   & 00031914076 & 05/19 & 10:34 & 10:58 & 1479 & 96414-01-16-06 & 05/19 & 05:15 & 05:36 & 832\\
 55706   & 00031914078 & 05/25 & 12:36 & 12:57 & 1245 & 96414-01-17-03 & 05/25 & 03:46 & 04:42 & 2688\\
 55709   & 00031914079 & 05/28 & 01:35 & 01:56 & 1220 & 96414-01-18-01 & 05/28 & 02:12 & 03:13 & 3152\\
 55712   & 00031914080 & 05/31 & 22:48 & 23:05 & 1055 & 96414-01-18-05 & 05/31 & 21:33 & 21:54 & 1280\\
 55715   & 00031914081 & 06/03 & 19:57 & 20:15 & 1101 & 96414-01-19-00 & 06/04 & 03:57 & 04:33 & 2160\\
 55718   & 00031914082 & 06/06 & 00:50 & 01:07 & 1059 & 96414-01-19-01 & 06/06 & 02:37 & 03:30 & 3088\\
 55721   & 00031914083 & 06/09 & 01:05 & 01:27 & 1322 & 96414-01-19-03 & 06/09 & 02:43 & 04:15 & 3120\\
 55724   & 00031914084 & 06/12 & 01:21 & 01:44 & 1377 & 96414-01-20-01 & 06/12 & 02:48 & 03:42 & 3184\\
 55746   & 00031914091 & 07/04 & 16:06 & 16:29 & 1330 & 96414-01-23-01 & 07/05 & 02:32 & 02:56 & 1024\\
 55749   & 00031914092 & 07/07 & 19:28 & 19:43 & 922 & 96414-01-24-00 & 07/08 & 00:57 & 02:56 & 3120\\
 55752   & 00031914093 & 07/10 & 00:35 & 00:36 & 66 & 96414-01-24-02 & 07/10 & 03:12 & 03:34 & 864\\
 55758   & 00031914095 & 07/16 & 12:02 & 12:18 & 974 & 96414-01-25-00 & 07/16 & 21:54 & 00:30 & 6288\\
 55774   & 00031914099 & 08/01 & 10:35 & 10:51 & 997 & 96414-01-27-04 & 08/01 & 21:57 & 00:26 & 6032\\
 55778   & 00031914100 & 08/05 & 18:34 & 18:49 & 948 & 96414-01-28-00 & 08/05 & 13:46 & 13:59 & 800\\
 55782   & 00031914101 & 08/09 & 22:04 & 22:21 & 974 & 96414-01-28-04 & 08/09 & 18:12 & 18:33 & 1280\\
 55785   & 00031914102 & 08/13 & 06:38 & 06:55 & 1021 & 96414-01-29-00 & 08/12 & 23:24 & 23:39 & 720\\
 55790   & 00031914103 & 08/17 & 00:13 & 00:29 & 943 & 96414-01-29-03 & 08/17 & 03:22 & 03:43 & 784\\
 55805   & 00031914107 & 09/02 & 03:13 & 05:01 & 6463 & 96414-01-31-05 & 09/01 & 20:35 & 21:16 & 2096\\
 55809   & 00031914108 & 09/06 & 00:20 & 08:36 & 29714 & 96414-01-32-01 & 09/05 & 19:03 & 19:27 & 1472\\
 55826   & 00031914112 & 09/22 & 06:40 & 06:57 & 974 & 96414-01-34-02 & 09/22 & 15:05 & 15:22 & 1024\\
 55832   & 00031914113 & 09/29 & 05:46 & 05:57 & 678 & 96414-01-35-02 & 09/28 & 23:07 & 23:40 & 1792\\
\hline

\end{tabular}
\end{center}
$^{*}$: Energy spectra are shown in figure \ref{fig:spectra} and the best-fit parameters are summarized table \ref{tab:spec_parameters}.\\
$^\dagger$: Date and time are in UT (Universal Time).\\
\end{table*}

%% file: spectable.tex
\begin{table*}
\renewcommand{\arraystretch}{0.4}

\caption{Best-fit model parameters of six sample epochs obtained by Swift/XRT and RXTE/PCA combined spectral analysis}
\label{tab:spec_parameters}
\begin{center}
\small
\begin{tabular}{lllllll}

\hline
\hline
\multicolumn{7}{c}{Observation-log Summary} \\ 
           & Obs\#1 & Obs\#2 & Obs\#3 & Obs\#4 & Obs\#5 & Obs\#6 \\
\hline
 Date(UT)  & 2011-01-13 & 2011-01-23 & 2011-02-08 & 2011-02-28 & 2011-03-20 & 2011-04-11 \\
 MJD       & 55574 & 55584 & 55600 & 55620 & 55640 & 55662 \\
\hline
\hline

\multicolumn{7}{c}{Model: {\tt wabs*cutoffpl}} \\ 

Parameter & Obs\#1 & Obs\#2 & Obs\#3 & Obs\#4 & Obs\#5 & Obs\#6 \\
\hline
$N_{\rm H}$ [10$^{21}$ cm$^{-2}$]& $0.63^{+0.06}_{-0.06}$ & $0.63^{+0.06}_{-0.06}$ & $0.65^{+0.08}_{-0.08}$ & $0.65^{+0.10}_{-0.10}$ & $0.34^{+0.12}_{-0.12}$ & $0.39^{+0.12}_{-0.12}$ \\
$\Gamma$& $0.45^{+0.03}_{-0.03}$ & $0.88^{+0.03}_{-0.03}$ & $0.91^{+0.04}_{-0.04}$ & $0.81^{+0.04}_{-0.04}$ & $0.54^{+0.05}_{-0.05}$ & $0.85^{+0.05}_{-0.05}$ \\
$E_{\rm cut}$ [keV]& $1.56^{+0.02}_{-0.02}$ & $2.19^{+0.03}_{-0.03}$ & $3.03^{+0.06}_{-0.06}$ & $3.51^{+0.09}_{-0.08}$ & $3.23^{+0.09}_{-0.09}$ & $4.52^{+0.17}_{-0.16}$ \\
$f_{\rm PCA/XRT}$& $0.89^{+0.01}_{-0.01}$ & $1.06^{+0.01}_{-0.01}$ & $1.11^{+0.02}_{-0.02}$ & $1.07^{+0.02}_{-0.02}$ & $1.12^{+0.02}_{-0.02}$ & $1.02^{+0.02}_{-0.02}$ \\
\hline
$F_{\rm CPL}$& 4.01 & 2.91 & 2.10 & 1.74 & 1.30 & 0.98 \\
$\chi^2_\nu$ (d.o.f.)& 1.32 (426) & 1.26 (389) & 1.17 (316) & 1.04 (309) & 0.94 (251) & 1.10 (187) \\

\hline
\hline
\multicolumn{7}{c}{Model: {\tt wabs*(diskBB+BB)}} \\ 
Parameter & Obs\#1 & Obs\#2 & Obs\#3 & Obs\#4 & Obs\#5 & Obs\#6 \\
\hline

$N_{\rm H}$ [10$^{21}$ cm$^{-2}$]& $0.54^{+0.04}_{-0.04}$ & $0.28^{+0.05}_{-0.05}$ & $0.21^{+0.06}_{-0.06}$ & $0.24^{+0.06}_{-0.06}$ & $0.24^{+0.07}_{-0.08}$ & $0.00^{+0.06}_{-}$ \\
$kT_{\rm in}$ [keV]& $0.96^{+0.02}_{-0.02}$ & $0.80^{+0.02}_{-0.02}$ & $0.99^{+0.03}_{-0.03}$ & $1.30^{+0.06}_{-0.05}$ & $1.69^{+0.11}_{-0.10}$ & $1.37^{+0.03}_{-0.06}$ \\
$R_{\rm in}\sqrt{\cos i}$ [km]& $13.9^{+0.6}_{-0.5}$ & $14.9^{+0.6}_{-0.6}$ & $7.9^{+0.4}_{-0.4}$ & $4.4^{+0.3}_{-0.3}$ & $2.5^{+0.2}_{-0.2}$ & $2.7^{+0.2}_{-0.2}$ \\
$kT_{\rm BB}$ [keV]& $1.32^{+0.05}_{-0.05}$ & $1.25^{+0.02}_{-0.02}$ & $1.61^{+0.03}_{-0.03}$ & $2.03^{+0.07}_{-0.06}$ & $2.49^{+0.36}_{-0.22}$ & $2.25^{+0.09}_{-0.07}$ \\
$R_{\rm BB, 10kpc}$ [km]& $3.7^{+0.6}_{-0.6}$ & $5.5^{+0.3}_{-0.3}$ & $3.0^{+0.2}_{-0.2}$ & $1.7^{+0.2}_{-0.2}$ & $0.8^{+0.3}_{-0.3}$ & $1.2^{+0.1}_{-0.1}$ \\
$f_{\rm PCA/XRT}$& $0.89^{+0.01}_{-0.01}$ & $1.07^{+0.01}_{-0.01}$ & $1.11^{+0.02}_{-0.02}$ & $1.05^{+0.02}_{-0.02}$ & $1.11^{+0.02}_{-0.02}$ & $1.01^{+0.02}_{-0.02}$ \\
\hline
$F_{\rm diskBB}$& 3.54 & 1.87 & 1.30 & 1.16 & 1.05 & 0.57 \\
$F_{\rm BB}$& 0.44 & 0.79 & 0.63 & 0.51 & 0.26 & 0.37 \\
$\chi^2_\nu$ (d.o.f.)& 1.25 (425) & 1.54 (388) & 1.46 (315) & 1.15 (308) & 0.90 (250) & 1.31 (186) \\

%

\hline
\hline
\multicolumn{7}{c}{Model: {\tt wabs*(BB+nthcomp)}} \\ 
Parameter & Obs\#1 & Obs\#2 & Obs\#3 & Obs\#4 & Obs\#5 & Obs\#6 \\

\hline

$N_{\rm H}$ [10$^{21}$ cm$^{-2}$]& $0.00^{+0.17}_{-}$ & $0.00^{+0.08}_{-}$ & $0.00^{+0.09}_{-}$ & $0.00^{+0.21}_{-}$ & $0.00^{+0.11}_{-}$ & $0.00^{+0.55}_{-}$ \\
$kT_{\rm BB}$ [keV]& $0.28^{+0.01}_{-0.01}$ & $0.28^{+0.01}_{-0.01}$ & $0.29^{+0.01}_{-0.01}$ & $0.29^{+0.01}_{-0.02}$ & $0.28^{+0.02}_{-0.02}$ & $0.27^{+0.02}_{-0.02}$ \\
$R_{\rm BB}$ [km]& $98^{+18}_{-3}$ & $95^{+7}_{-3}$ & $67^{+5}_{-3}$ & $52^{+10}_{-3}$ & $43^{+5}_{-3}$ & $43^{+25}_{-3}$ \\
$\Gamma$& $2.98^{+0.23}_{-0.18}$ & $2.51^{+0.16}_{-0.16}$ & $2.26^{+0.07}_{-0.10}$ & $2.17^{+0.09}_{-0.08}$ & $2.17^{+0.16}_{-0.12}$ & $1.98^{+0.09}_{-0.07}$ \\
$kT_{\rm e}$ [keV]& $1.52^{+0.12}_{-0.09}$ & $1.63^{+0.12}_{-0.13}$ & $2.07^{+0.13}_{-0.06}$ & $2.50^{+0.17}_{-0.13}$ & $2.57^{+0.29}_{-0.19}$ & $2.79^{+0.22}_{-0.16}$ \\
$kT_{\rm seed}$ [keV]& $0.58^{+0.03}_{-0.03}$ & $0.60^{+0.04}_{-0.05}$ & $0.64^{+0.05}_{-0.02}$ & $0.67^{+0.05}_{-0.05}$ & $0.73^{+0.06}_{-0.09}$ & $0.67^{+0.06}_{-0.09}$ \\
$\tau$& $8.7^{+1.0}_{-1.1}$ & $10.3^{+1.4}_{-1.1}$ & $10.3^{+0.8}_{-0.8}$ & $9.8^{+0.8}_{-0.8}$ & $9.7^{+1.2}_{-1.4}$ & $10.5^{+0.9}_{-1.0}$ \\
$R_{\rm seed}$ [km]& $44^{+3}_{-3}$ & $32^{+4}_{-3}$ & $24^{+1}_{-3}$ & $21^{+3}_{-2}$ & $16^{+3}_{-2}$ & $15^{+4}_{-2}$ \\
$f_{\rm PCA/XRT}$& $0.88^{+0.01}_{-0.01}$ & $1.03^{+0.01}_{-0.01}$ & $1.09^{+0.02}_{-0.02}$ & $1.04^{+0.02}_{-0.02}$ & $1.10^{+0.02}_{-0.02}$ & $1.00^{+0.02}_{-0.02}$ \\
\hline
$F_{\rm BB}$& 1.16 & 1.14 & 0.66 & 0.38 & 0.24 & 0.20 \\
$F_{\rm nthcomp}$& 2.92 & 1.85 & 1.48 & 1.40 & 1.14 & 0.81 \\
$\chi^2_\nu$ (d.o.f.)& 1.19 (423) & 1.14 (386) & 1.10 (313) & 1.00 (306) & 0.89 (248) & 1.02 (184) \\

\hline
\hline
\multicolumn{7}{c}{Model: {\tt wabs*(diskBB+nthcomp)}} \\ 
Parameter & Obs\#1 & Obs\#2 & Obs\#3 & Obs\#4 & Obs\#5 & Obs\#6 \\

\hline

$N_{\rm H}$ [10$^{21}$ cm$^{-2}$]& $0.93^{+0.16}_{-0.15}$ & $0.79^{+0.11}_{-0.11}$ & $0.76^{+0.16}_{-0.15}$ & $0.85^{+0.23}_{-0.23}$ & $0.53^{+0.24}_{-0.22}$ & $0.89^{+0.35}_{-0.32}$ \\
$kT_{\rm in}$ [keV]& $0.34^{+0.03}_{-0.03}$ & $0.42^{+0.03}_{-0.03}$ & $0.45^{+0.05}_{-0.04}$ & $0.40^{+0.07}_{-0.05}$ & $0.50^{+0.12}_{-0.08}$ & $0.37^{+0.07}_{-0.05}$ \\
$R_{\rm in}\sqrt{\cos i}$ [km]& $71^{+16}_{-13}$ & $44^{+7}_{-6}$ & $29^{+7}_{-6}$ & $27^{+9}_{-7}$ & $14^{+6}_{-5}$ & $24^{+11}_{-8}$ \\
$\Gamma$& $2.75^{+0.04}_{-0.03}$ & $2.72^{+0.05}_{-0.05}$ & $2.33^{+0.04}_{-0.04}$ & $2.19^{+0.03}_{-0.03}$ & $2.31^{+0.06}_{-0.05}$ & $2.02^{+0.03}_{-0.03}$ \\
$kT_{\rm e}$ [keV]& $1.41^{+0.10}_{-0.07}$ & $1.72^{+0.19}_{-0.13}$ & $2.12^{+0.15}_{-0.12}$ & $2.51^{+0.19}_{-0.15}$ & $2.76^{+0.43}_{-0.27}$ & $2.88^{+0.28}_{-0.20}$ \\
$kT_{\rm seed}$ [keV]& $0.55^{+0.02}_{-0.02}$ & $0.67^{+0.03}_{-0.03}$ & $0.70^{+0.05}_{-0.04}$ & $0.69^{+0.05}_{-0.04}$ & $0.82^{+0.07}_{-0.05}$ & $0.71^{+0.06}_{-0.05}$ \\
$\tau$ & $10.0^{+0.5}_{-0.6}$ & $9.1^{+0.6}_{-0.7}$ & $9.8^{+0.5}_{-0.6}$ & $9.7^{+0.5}_{-0.6}$ & $8.5^{+0.8}_{-1.0}$ & $10.0^{+0.6}_{-0.7}$ \\
$R_{\rm seed}$ [km]& $51^{+2}_{-2}$ & $25^{+2}_{-2}$ & $20^{+2}_{-2}$ & $20^{+2}_{-2}$ & $13^{+1}_{-2}$ & $14^{+2}_{-1}$ \\
$f_{\rm PCA/XRT}$& $0.88^{+0.01}_{-0.01}$ & $1.03^{+0.01}_{-0.01}$ & $1.10^{+0.02}_{-0.02}$ & $1.05^{+0.02}_{-0.02}$ & $1.10^{+0.02}_{-0.02}$ & $1.00^{+0.02}_{-0.02}$ \\
\hline
$F_{\rm diskBB}$& 1.32 & 1.25 & 0.70 & 0.41 & 0.26 & 0.23 \\
$F_{\rm nthcomp}$& 3.02 & 1.71 & 1.40 & 1.38 & 1.09 & 0.81 \\
$\chi^2_\nu$ (d.o.f.)& 1.19 (423) & 1.09 (386) & 1.10 (313) & 1.00 (306) & 0.88 (248) & 1.01 (184) \\

\hline

\end{tabular}
\end{center}
\begin{enumerate}[(i)]
\scriptsize

\item All errors represent the 90\% confidence limits of statistical
  uncertainty for a single parameter of interest.
\item $R_{\rm BB}$ is a {\tt BB}-model parameter representing
  the source radius if the source distance is 10 kpc.
\item $R_{\rm in} \sqrt{\cos i}$ is a {\tt diskBB}-model
  parameter related to the inner radius of the accretion disk $R_{\rm in}$
  and the disk inclination $i$ 
  in the assumed  10 kpc source distance. 
\item $R_{\rm seed}$ is a radius of blackbody seed photons in 
  the thermally Compotonized contimuum model ({\tt nthcomp})
  in the assumed 10 kpc source distance.
\item $F_{\rm CPL}$, $F_{\rm BB}$, $F_{\rm DBB}$, $F_{\rm nthcomp}$
  represent absorption-corrected fluxes of each continuum model of
  {\tt cutoffpl}, {\tt BB}, {\tt diskbb}, and {\tt nthcomp},
  respectively, in units of 10$^{-9}$ erg cm$^{-2}$ s$^{-1}$ in
  0.1-100 keV band.\\
\end{enumerate}
\end{table*}